\documentclass{aa}

\usepackage{graphicx}
\usepackage{txfonts}
\usepackage[colorlinks=true,
            urlcolor=blue,
            linkcolor=blue,
            citecolor=blue]
            {hyperref}
\usepackage{tabu}
\usepackage{ulem}

\newcommand{\Ha}{H$\alpha$}
\newcommand{\Hb}{H$\beta$}
\newcommand{\Hg}{H$\gamma$}

\newcommand{\ngc}{NGC\,1566{}}

\newcommand{\swift}{\textit{Swift}{}}

\newcommand{\xmm}{\textit{XMM}-Newton{}}

\newcommand{\nustar}{NuSTAR{}}

\newcommand{\arc}{\arcsec\hspace*{-1ex}.\hspace*{0.3ex}}

\newcommand{\mcc}[1]{\multicolumn{1}{c}{#1}}

\newcommand{\kms}{km\,s$^{-1}$}

\begin{document} 

   \title{The transient event in \ngc{} from 2017 to 2019\\ I. An eccentric accretion disk and a turbulent, disk-dominated broad-line region unveiled by double-peaked \ion{Ca}{ii} and \ion{O}{i} lines\thanks{Based on observations made with the Southern African Large Telescope (SALT).}
   }

      \author{M.W. Ochmann \inst{1,2},
           W.~Kollatschny \inst{1},
           M.A.~Probst \inst{1},
           E.~Romero-Colmenero \inst{3,4},
           D.A.H.~Buckley \inst{3,4,5,6},
           D.~Chelouche \inst{7,8},
           R.~Chini \inst{2,9,10},
           D.~Grupe \inst{11},
           M.~Haas \inst{2},
           S.~Kaspi \inst{12},
           S.~Komossa \inst{13},      
           M.L.~Parker \inst{14},
           M.~Santos-Lleo \inst{15},
           N.~Schartel \inst{15},
           and P.~Famula \inst{1}
           }

   \institute{
          Institut f\"ur Astrophysik und Geophysik, Universit\"at G\"ottingen,
          Friedrich-Hund Platz 1, 37077 G\"ottingen, Germany \\
          \email{martin.ochmann@uni-goettingen.de}
          \and
          Ruhr University Bochum, Faculty of Physics and Astronomy, Astronomical Institute (AIRUB), 44780 Bochum, Germany
          \and
          South African Astronomical Observatory, P.O. Box 9, Observatory Road, Observatory 7935, Cape Town, South Africa
          \and
          Southern African Large Telescope, P.O. Box 9, Observatory, 7935, South Africa
          \and
          Department of Astronomy, University of Cape Town, Private Bag X3, Rondebosch 7701, South Africa
          \and
          Department of Physics, University of the Free State, P.O. Box 339, Bloemfontein 9300, South Africa
          \and
          Department of Physics, Faculty of Natural Sciences, University of Haifa, Haifa 3498838, Israel
          \and
          Haifa Research Center for Theoretical Physics and Astrophysics, University of Haifa, Haifa 3498838, Israel
          \and
          Nicolaus Copernicus Astronomical Center, Polish Academy of Sciences, Bartycka 18, 00-716 Warszawa, Poland
          \and
          Universidad Católica del Norte, Instituto de Astronomía, Avenida Angamos 0610, Antofagasta, Chile
          \and
          Department of Physics, Geology, and Engineering Technology, Northern Kentucky University, 1 Nunn Drive, Highland Heights, KY 41099, USA
          \and
          School of Physics \& Astronomy and the Wise Observatory, The Raymond and Beverly Sackler Faculty of Exact Sciences, Tel-Aviv University, Tel-Aviv 69978, Israel
          \and
          Max-Planck-Institut für Radioastronomie, Auf dem Hügel 69, D-53121 Bonn, Germany
          \and
          Institute of Astronomy, Madingley Road, Cambridge CB3 0HA, UK
          \and
          European Space Agency (ESA), European Space Astronomy Centre (ESAC), Villanueva de la Cañada, E-28691 Madrid, Spain
          }

  \date{Received 2023 November 10; accepted 2024 February 16}

 \abstract{
\ngc{} is a local face-on Seyfert galaxy and is known for exhibiting recurrent outbursts that are accompanied by changes in spectral type. The most recent transient event occurred from 2017 to 2019 and was reported to be accompanied by a change in Seyfert classification from Seyfert 1.8 to Seyfert 1.2.
}
{
We aim to study the transient event in detail by analyzing the variations in the optical broad-line profiles. In particular, we intend to determine the structure and kinematics of the broad-line region.
}
{
We analyzed data from an optical spectroscopic variability campaign of \ngc{} taken with the 9.2\,m Southern African Large Telescope (SALT) between July 2018 and October 2019 triggered by the detection of hard X-ray emission in June 2018. We supplemented this data set with optical to near-infrared (NIR) spectroscopic archival data taken by VLT/MUSE in September 2015 and October 2017, and investigated the emission from different line species during the event.

}
{
\ngc{} exhibits pronounced spectral changes during the transient event. We observe the emergence and fading of a strong power-law-like blue continuum as well as strong variations in the Balmer, \ion{He}{i}, and \ion{He}{ii} lines and the coronal lines [\ion{Fe}{vii}], [\ion{Fe}{X}], and [\ion{Fe}{XI}]. Moreover, we detect broad double-peaked emission line profiles of \ion{O}{i}\,$\lambda8446$ and the \ion{Ca}{ii}$\,\lambda\lambda8498,8542,8662$ triplet. This is the first time that genuine double-peaked \ion{O}{i}\,$\lambda8446$ and \ion{Ca}{ii}$\,\lambda\lambda8498,8542,8662$ emission in AGN is reported in the literature.  All broad lines show a clear redward asymmetry with respect to their central wavelength and  we find indications for a significant blueward drift of the total line profiles during the transient event. The profiles and the FWHM of the Balmer lines remain largely constant during all observations. We show that the double-peaked emission line profiles are well approximated by emission from a low-inclination, relativistic eccentric accretion disk, and that single-peaked profiles can be obtained by broadening due to scale-height-dependent turbulence. Small-scale features in the \ion{O}{i} and \ion{Ca}{ii} lines suggest the presence of inhomogeneities in the broad-line region.
}
{
We conclude that the broad-line region in \ngc{} is dominated by the kinematics of a relativistic eccentric accretion disk. The broad-line region can be modeled to be vertically stratified with respect to scale-height turbulence with \ion{O}{i} and \ion{Ca}{ii} being emitted close to the disk in a region with high (column) density, while the Balmer and helium lines are emitted at greater scale height above the disk.  The observed blueward drift might be attributed to a low-optical-depth wind launched during the transient event. Except for this wind, the observed kinematics of the broad-line region remain largely unchanged during the transient event.
} 


  \titlerunning{A disk-dominated BLR in \ngc{}}
  \authorrunning{M.W. Ochmann et al.}

  \maketitle
%
%

%
\section{Introduction}\label{sec:introduction}
%
Variability is widespread in active galactic nuclei (AGN). It occurs in all spectral bands and on typical timescales of hours to weeks or even years. Generally, the variability of AGN is assumed to be stochastic in nature and has been used with great success in the last $\sim$ 30 years to identify and map the innermost AGN structures ---namely the accretion disk (AD), the broad-line region (BLR), and the dusty torus (TOR)--- using methods such as reverberation mapping \citep[RM;][]{blandford82}. RM traces the lagging emissive response of individual AGN emission lines to the time-varying ionizing continuum radiation from the central source close to the supermassive black hole (SMBH). As regular and densely sampled observations of the X-ray/UV ionizing continuum are difficult to acquire, an optical continuum is often used as a proxy for the ionizing radiation. Typical optical continuum and line variability for an object (on timescales of months) can span a wide range, from only a few percent up to a few dozen percent \citep[e.g.,][]{ulrich97}. Aside from these typical temporal variations of the continuum and emission lines, studies have shown that the variability behavior of individual objects can differ from one epoch to another, and apparent variations of the BLR responsivity have been reported \citep[e.g.,][]{hu20,gaskell21}, sometimes over timescales comparable to or even shorter than the expected dynamical timescales \citep{derosa18}.

In addition to the overall stochastic variability behavior of AGN, transient events such as changing-look (CL) transitions have increasingly gained attention over recent years.  Originally, the term ``changing-look'' was used to describe Compton-thick AGN becoming Compton-thin and vice versa \citep[e.g.,][]{guainazzi02, matt03}. In analogy, optical CL AGN are characterized by their change of spectral classification, switching between Sy 1 and Sy 2 and associated subtypes\footnote{Sy 1 galaxies show Balmer lines that are wider than the forbidden lines, while Sy 2 galaxies show Balmer lines that are about the same width as the forbidden lines \citep{khachikian74}. The types Sy 1.2, Sy 1.5, Sy 1.8, and Sy 1.9 galaxies are subtypes, with numerically larger Seyfert classes showing weaker broad emission with respect to the narrow lines \citep{osterbrock77, osterbrock81, winkler92}.}. These transitions happen over timescales of months to years and are often accompanied by significant flux changes on the order of several magnitudes  \citep[e.g.,][]{macleod16,graham20,green22}. 
The observed change between Seyfert types during CL events does not challenge the general validity of the unified AGN model, according to which the source classification is mainly due to the orientation with respect to the observer \citep{antonucci93}. Rather, the huge change in continuum flux on short timescales and the resulting apparent changes in BLR kinematics provide a unique opportunity to refine our understanding of BLR structure. According to the locally optimized cloud model \citep[LOC model][]{baldwin95}, the continuum luminosity determines which parts of the BLR ---near or far from the continuum source--- are visible. If the BLR is not scale invariant, then a transient event inevitably leads to obvious changes in BLR kinematics and Seyfert subtype. However, to date, the implications of the LOC model have not been adequately addressed in CL-AGN research.

The typical CL transition timescales cannot be explained by viscous radial inflow, a circumstance that is known as the ``viscosity crisis'' \citep[][and references therein]{lawrence18}. Currently, several explanations for the CL phenomenon are discussed, including tidal disruption events (TDEs), strong variations in the accretion flow, microlensing caused by an intervening object, or sudden changes in obscuration. However, at least for some cases, the behavior of the observed post-CL light curves disfavors discrete events as the cause of the CL phenomenon \citep[e.g.,][]{runnoe16,zetzl18}. In general, the similarities between TDEs and some observed CL events require clearly defined distinction criteria and, in turn, extensive observational data for each event \citep[e.g.,][]{zabludoff21,komossa22}.

Other possible mechanisms that are discussed include accretion disk instabilities \citep{nicastro03}, magneto-rotational instabilities \cite[e.g.,][]{ross18}, radiation pressure instabilities \citep{sniegowska23}, magnetically elevated accretion \citep[e.g.,][]{dexter19}, accretion state transitions \citep[e.g.,][]{noda18}, and interactions between binaries of SMBHs \citep{wang20}. In addition, the phenomenon of periodicities in AGN light curves \citep[e.g.,][and references therein]{bon17} and repeat CL events \citep[e.g.,][]{sniegowska20} has gained more attention from the scientific community in recent years.

Until recently, CL events were thought to be rather rare. However, the evidence for CL events being much more common has been growing in recent years. To date, a few dozen CL AGN have been identified. Early detections include \ngc{} \citep{pastoriza70}, NGC\,3515 \citep{collin-souffrin73}, NGC\,4151 \citep{penston84}, and Fairall\,9 \citep{kollatschny85}. More recent detections are, for example, NGC\,2617 \citep{shappee14}, Mrk\,590 \citep{denney14}, HE\,1136-2304 \citep{parker16, zetzl18, kollatschny18}, WISE\,J1052+1519 \citep{stern18}, and 1ES\,1927+654 \citep{trakhtenbrot19b}. However, only a few of them ---most notably 1ES\,1927+654--- have been studied spectroscopically in greater detail in temporal proximity to the transient event. This lack of high-quality data is a significant hinderance to endeavors to understand the CL phenomenon.

\ngc{} ($\alpha_{2000}=04^\text{h}20^\text{m}00.42^\text{s}, \delta_{2000} = -54^{\circ}56^{\prime}16.1^{\prime \prime}$) is a local ($z=0.00502$) face-on Seyfert galaxy\footnote{\url{https://ned.ipac.caltech.edu/}} and is known for exhibiting recurrent outbursts accompanied by changes in spectral type \citep[][]{shobbrook66,pastoriza70,alloin85,alloin86}.  The most recent transient event occurred from 2017 to 2019, and an accompanying CL event (a change in Seyfert classification from Sy 1.8 to Sy 1.2; that is, from showing only weak broad emission in \Hb{} and \Ha{} to showing stronger broad \Hb{} emission) was reported by \citet{oknyansky19, oknyansky20}. Optical (post-)outburst spectra from 2018 were presented by \citet{oknyansky19, oknyansky20} and \citet{ochmann20}. The flux and spectral variations were the strongest changes observed since 1962, when \ngc{} exhibited similarly strong broad-line emission \citep{shobbrook66, pastoriza70}. The object already started to brighten significantly in September 2017 \citep{dai18} and reached its peak optical flux in July 2018. A thorough historical overview of the variations in \ngc{} from the 1960s until today can be found in \citet{oknyansky20}. 

Here, we present first results of a multiwavelength campaign of \ngc{} during its transient event from 2017 to 2019.  Observations with SALT, \xmm{}, \nustar,{} and \swift{} were triggered by the detection of hard X-ray emission with Integral in June 2018 \citep{ducci18}. This led to a dense multiwavelength campaign with a duration of $\sim 850$ days and follow-up observations in 2019/2020. The first \xmm{}, \nustar,{} and \swift{} observations \citep{parker19} revealed the rapid increase in X-rays and the presence of a typical Seyfert-1-type X-ray spectrum in outburst, along with the formation of an X-ray wind at $v \sim 500$\,\kms{}. In 2023 October, \ngc{} was found to be in a low state in all \swift{} bands \citep{xu23}. Further results of the observations with \xmm{}, \nustar,{} and \swift{} will be presented in future publications. Our observations are supplemented by archival observations with VLT/MUSE, and cover the optical and NIR wavelength range ($\sim 4300$\,\AA{} $- \sim 9300$\,\AA{}) at epochs directly before, during, and after the transient event. The observations in detail reveal drastic changes in the line emission and nonstellar continuum. The present paper is structured as follows. In Sect.\,\ref{sec:observations}, we describe the observations and the data reduction. In Sect.\,\ref{sec:results}, we present the analysis of the spectroscopic observations. We discuss the results in Sect.\,\ref{sec:discussion} and summarize them in Sect.\,\ref{sec:conclusions}. Throughout this paper, we assume a $\Lambda$ cold dark matter cosmology with a Hubble constant of $H_0$~=~73~km s$^{-1}$ Mpc$^{-1}$, $\Omega_{\Lambda}$~=~0.73, and $\Omega_{\rm M}$~=~0.27.

%
\section{Observations and data reduction}\label{sec:observations}
%

%
\subsection{Optical spectroscopy with SALT}\label{sec:SALT_observations}
%
We acquired optical long-slit spectra of \ngc{} with the Southern African Large Telescope \citep[SALT;][]{buckley06} at 2 epochs shortly after the detection of hard X-ray emission in June 2018, and one follow-up spectrum in September 2019. The observations have proposal codes 2018-1-DDT-004, 2018-1-DDT-008 (PI: Kollatschny) and 2018-2-LSP-001 (PI: Buckley). The log of the spectroscopic observations is given in Table\,\ref{tab:spectroscopy_log}. In addition to the galaxy spectra, we took necessary calibration images (flat-fields, Xe arc frames). All observations were taken under identical instrumental conditions with the help of the Robert Stobie Spectrograph \citep[RSS;][]{kobulnicky03} using the PG0900 grating and a 2x2  spectroscopic binning. To minimize differential refraction, the slit width was fixed to 2\arc0 projected onto the sky at an optimized projection angle. For the extraction of the spectra, we used a square aperture of 2\arc0 $\times$ 2\arc0. We covered the wavelength range from 4210 to 7247\,\AA{} with a spectral resolution of $\sim 6.7$\,\AA{}. This corresponds to object rest-frame wavelengths of 4189 to 7211\,\AA{}. Two gaps in the spectra are caused by gaps between the three CCDs of the spectrograph. They range from 5219 to 5274\,\AA{} and 6262 to 6315\,\AA{} (5193 to 5248\,\AA{} and 6231 to 6283\,\AA{} in the rest-frame), respectively. For all observations, we used the same instrumental setup as well as the same standard star (LTT\,4364) for flux calibration, and performed standard reduction procedures using \textsc{iraf} packages. In order to account for small spectral shifts ($\lesssim 0.5$\,\AA{}) in the wavelength calibration between spectra, we performed a wavelength intercalibration with respect to the MUSE spectra. This was done separately for the \Hb{} and \Ha{} line using the narrow lines [\ion{O}{iii}]\,$\lambda\lambda4959,5007$ and [\ion{S}{ii}]\,$\lambda\lambda6716,6731$, respectively.

In addition to our three observations, we utilize one additional SALT observation of \ngc{} from the SALT archive\footnote{\url{https://ssda.saao.ac.za/}} for our variability study. This spectrum was taken on 2018 July 30 as part of the proposal 2018-1-SCI-029 (PI: Marchetti) by the RSS using the PG0900 grating, a 1\arc5 slit and 2x4 binning. This setup covered the wavelength range from 4920 to 7922\,\AA{} with a spectral resolution of $\sim 5.7$\,\AA{}. This corresponds to object rest-frame wavelengths of 4895 to 7882\,\AA. We followed the same reduction steps as for the other SALT observations, employing calibration images with matching instrumental setup. In particular, we used the same standard star LTT\,4364 for flux calibration of the spectrum. The signal-to-noise ratio (S/N) in the continuum range $(5100 \pm 20)$\,\AA{} (rest-frame) is $\sim 110$ compared to S/N $\sim 190$ in the 2\arc0 aperture SALT spectrum from 2018 July 20.

All spectra were corrected for Galactic reddening applying the extinction curve of \citet{cardelli89} and using a ratio $R$ of absolute extinction $A(V)$ to $E_{\rm B-V}=0.0079$ \citep{schlafly11} of 3.1, and calibrated to the same [\ion{O}{iii}]\,$\lambda5007$ flux of $(102 \pm 2)\, \times$\, 10$^{-15}$ ergs s$^{-1}$ cm$^{-2}$ (see \ref{sec:MUSE_observations}) in the optical regime. We also corrected for slightly different background flux contributions between observations using an intercalibration to a campaign (Ochmann et al., in prep.) with the UV-Optical Telescope \citep[UVOT;][]{roming05} of \swift{}. The differing background flux contributions arise due to differing observing conditions between observations and the large spatial extent of \ngc{} in the slit.

\begin{table}[ht!]
\caption{Log of spectroscopic observations of \ngc{} before, during and after the transient event in 2018.}
\centering
\resizebox{0.48\textwidth}{!}{
    \begin{tabular}{llcrcl}
        \hline \hline
        \noalign{\smallskip}
        ID & Mod. JD & UT Date & Exp. Time   & Seeing    & Instr.    \\
           &         &         &  \mcc{[s]}   & {\small FWHM}      &           \\
        \noalign{\smallskip}
        \hline 
        \noalign{\smallskip}
        1 & 57289.23       &   2015-09-24  &   2080&   2\arc25         & MUSE\tablefootmark{a}  \\
        2 & 58049.20       &   2017-10-23  &   3600&   0\arc89         & MUSE\tablefootmark{b}  \\
        3 & 58319.17            &       2018-07-20      &       600     &   1\arc6\,\,\,    & RSS   \\
        4 & 58329.16       &   2018-07-30  &   100 &   -\,\,\,         & RSS   \\
        5 & 58395.96            &       2018-10-04      &       600     &       1\arc9\,\,\,    & RSS        \\
        6 & 58735.03            &       2019-09-09      &       600     &       1\arc5\,\,\,    & RSS        \\

        \hline 
    \end{tabular}}
\tablefoot{\tablefootmark{a}{Obtained in no-AO wide field mode.} \tablefootmark{b}{Obtained in AO wide field mode.}}
\label{tab:spectroscopy_log}
\end{table}
%

%
\subsection{Optical and NIR spectroscopy with MUSE}\label{sec:MUSE_observations}
%

\ngc{} was observed with VLT/MUSE \citep[Multi Unit Spectroscopic Explorer;][]{bacon10, bacon14} IFU spectrograph as part of the ESO programs 096.D-0263 (PI: J. Lyman) and 0100.B-0116 (PI: C.M. Carollo) on 2015 September 24 and 2017 October 23, respectively. The former observation was carried out in the no-AO wide field mode (WFM), that is, with natural seeing and FoV of $1'\times1'$, while the latter was performed in the AO WFM making use of adaptive optics. MUSE covers the optical and NIR wavelength range between $\sim 4700$\,\AA\ and $9300$\,\AA\ at a spectral resolution of $\sim 2.5$\,\AA. The spectra are sampled at 1.25\,\AA\ in dispersion direction and at 0\arc2 in spatial direction. The seeing and exposure time of the observations are given in Table~\ref{tab:spectroscopy_log}.

\begin{figure}[h!]
    \centering
    \includegraphics[width=0.47\textwidth,angle=0]{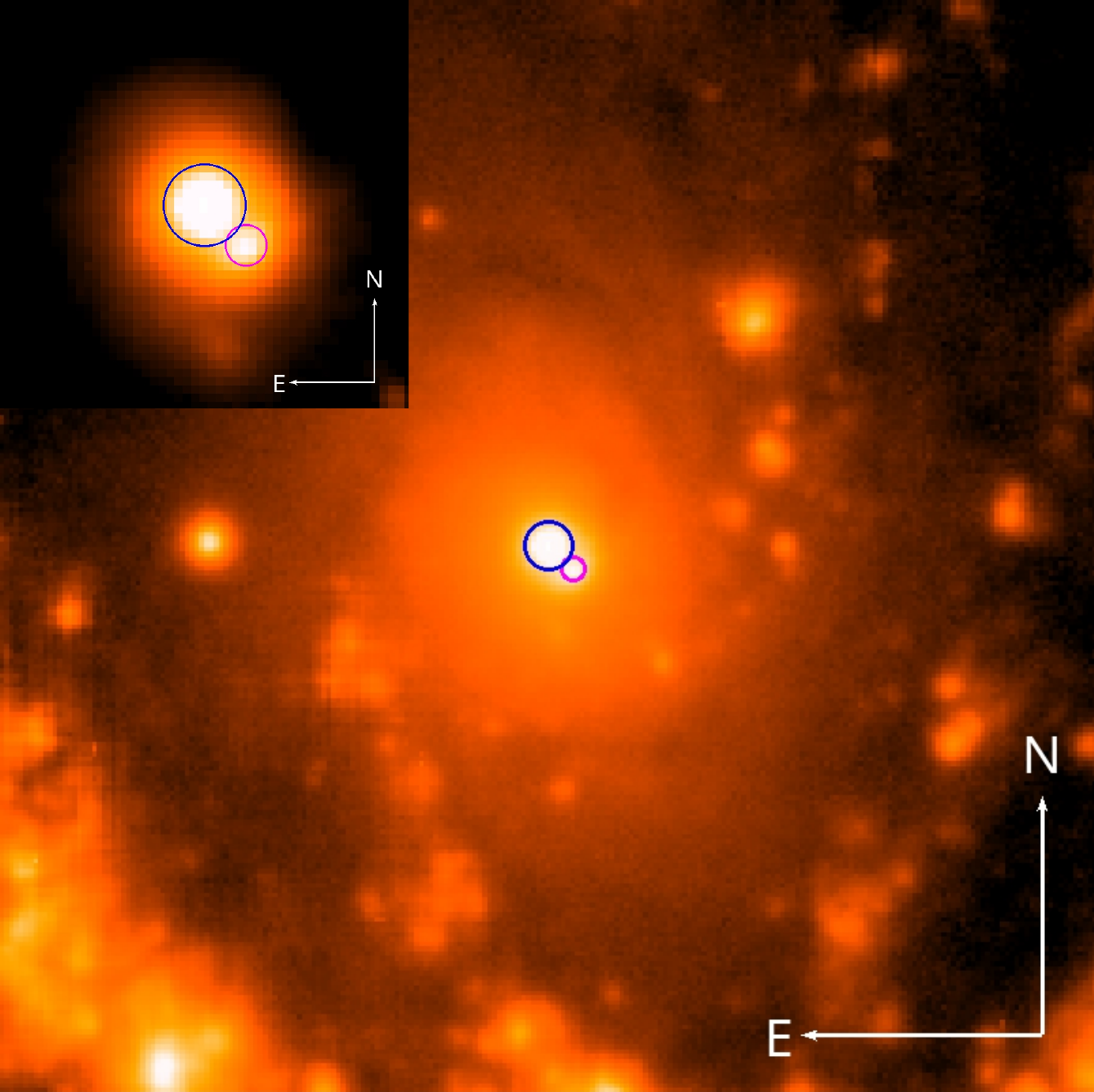}
    \caption{Image (45\arc 0 x 45\arc 0) of the central region of \ngc{} at 6597\,\AA{} (observed frame) taken by MUSE on 2017 October 23. Each arm of the compass is 10\arc 0 in length. Spectra of the nucleus and the \ion{H}{ii} region were extracted using apertures of 1\arc0 and 0\arc 5 in radius, respectively, indicated by the blue and magenta circle. The inlay shows the zoomed-in nuclear region (10\arc 0 x 10\arc 0). Each arm of the compass is 2\arc 0 in length. The color scale is logarithmic in order to enhance weaker emission features.}
    \label{fig:apertures_ngc1566}
\end{figure}

The data were reduced using the MUSE pipeline development version 1.6.1 and 2.2 \citep{weilbacher12, weilbacher14} for the observation from 2015 September 24 and 2017 October 23, respectively. This reduction includes the usual steps of bias subtraction, flat-fielding using a lamp-flat, wavelength calibration and twilight sky correction. Every data cube is the product of four combined raw science images. We extracted spectra of \ngc's nucleus and the \ion{H}{II} region detected by \citet{silva18} using circular apertures of 1\arc0 and 0\arc5 radius, respectively. The apertures were chosen such that they are centered on the respective region, comprise the bulk of the emission, and have minimal overlap. A zoomed-in region of the data cube from 2017 October 23 centered on the nucleus is shown in Fig.~\ref{fig:apertures_ngc1566}. The apertures are indicated by a blue and magenta circle, respectively.

In the following, we use the AO wide field mode MUSE spectrum from 2017 October 23 as a reference spectrum for all spectroscopic observations. Therefore, we calibrated all spectra to the same absolute [\ion{O}{iii}]\,$\lambda$5007 flux of $(102 \pm 2)\, \times$\, 10$^{-15}$ ergs s$^{-1}$ cm$^{-2}$.  This value is in agreement with results of \citet{kriss91}, who measured an [\ion{O}{iii}]\,$\lambda$5007 flux of ($101.62 \pm 7.32) \, \times $\, 10$^{-15}$ ergs s$^{-1}$ cm$^{-2}$ in a HST/FOS spectrum obtained with an aperture of 0\arc3 on 1991 February 8. This indicates that the bulk of the [\ion{O}{iii}]\,$\lambda$5007 emission close to the nucleus stems from a confined region with a size $\lesssim$ 0\arc3, which translates to $\lesssim 30$\,pc using the Cosmology Calculator of \citet{wright06}.

%

%
\section{Results}\label{sec:results}
%

%
\subsection{Optical spectral observations}\label{sec:optical_spectral_variations_results}
\begin{figure*}[!htp]
    \centering
    \includegraphics[width=1\textwidth,angle=0]{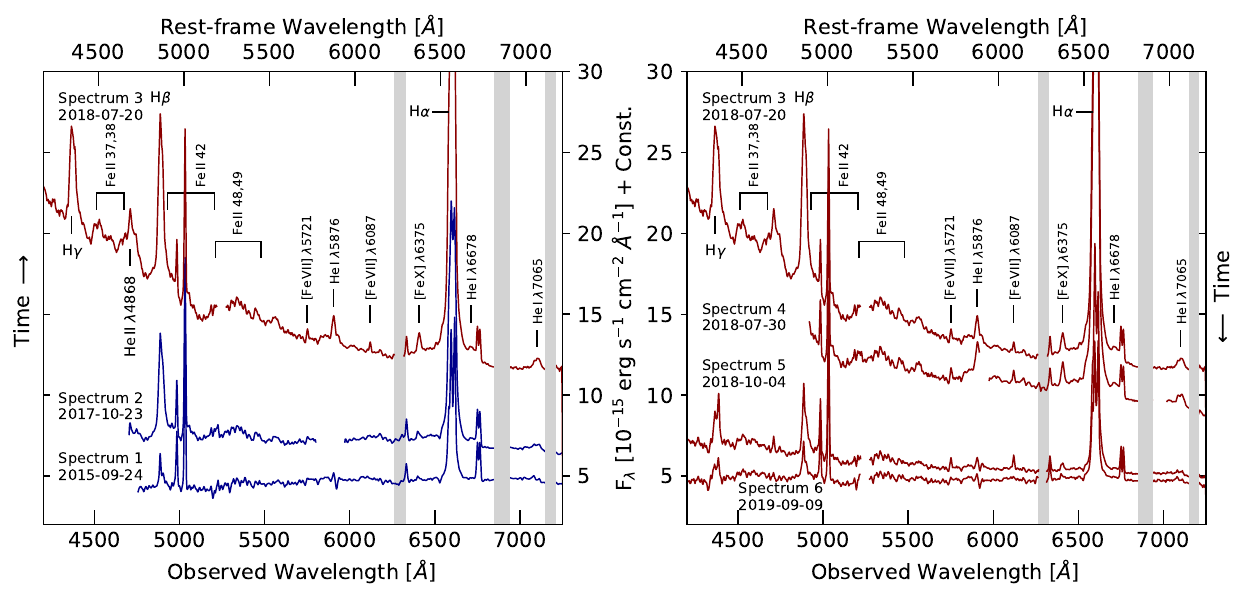}
    \caption{All optical spectra obtained before, during, and after the transient event in \ngc{}.  MUSE and SALT spectra are shown in blue and red, respectively. The left panel shows the spectra obtained during the rising phase, including the optical spectrum from 2018 July 20, while the right panel shows the spectra obtained during the declining phase, again including the optical high-state spectrum from 2018 July 20 for reference. The SALT spectrum from 2018 July 30 is shifted by $-2 \times 10^{-15}$ ergs s$^{-1}$ cm$^{-2}$\AA{}$^{-1}$ for clarity. 
    For each spectrum, we give the ID as well as the UT date of the observation. The most prominent telluric absorption bands are flagged (gray).}
    \label{fig:all_spectra}
\end{figure*}

%
We present all reduced optical spectra obtained before, during, and after the transient event in \ngc{} in Fig.~\ref{fig:all_spectra}. For each spectrum, we give a chronologically sorted ID, the UT date and the time difference in days with respect to 2018 July 02 ($t_0 = 58301.44$\,MJD), when \ngc{} reached its peak optical flux in the ASAS-SN\footnote{\url{http://www.astronomy.ohio-state.edu/~assassin}} \citep[All-Sky  Automated  Survey for SuperNovae;][]{shappee14, kochanek17, jayasinghe19} V-band and g-band light curves.

Spectrum 1 was obtained on 2015 September 24, and therefore $\sim 700$\,days before \citet{dai18} reported a brightening of \ngc{} in September 2017, and $1012$\,days before the transient event reached its peak. Spectrum 2\footnote{ We note that the narrow component of \Ha{} in Spectrum 2 obtained with MUSE is affected by overexposure, which considerably reduces the measured flux of \Ha{}$_{\rm narrow}$.} was obtained on 2017 October 23, $252$\,days before peak flux. This spectrum sees the emergence of a nearly linear continuum across the optical band, accompanied by the appearance of strong \ion{Fe}{ii} multiplet emission of the transitions 42 ($\sim 4910 - 5180$\,\AA{}), 48 and 49 ($\sim 5185 - 5450$\,\AA{}), and weak coronal line emission of [\ion{Fe}{vii}]$\,\lambda\lambda$5721,6087 and [\ion{Fe}{x}]$\,\lambda$6375 as well as weak emission of \ion{He}{i}$\,\lambda\lambda6678,7065$.

Spectra 3 and 4 were obtained on 2018 July 20 and July 30, and therefore $18$ and $28$\,days after peak flux. Spectrum 3 has already been presented by \citet{ochmann20}, however, it had not been intercalibrated to the other spectra of the campaign (see \ref{sec:SALT_observations}). To our knowledge, these two high-state spectra presented here are the optical spectra closest to the transient peak presented in the literature so far \citep[see][]{oknyansky19,oknyansky20}. The two spectra are qualitatively identical and show a strong, power-law-like blue continuum, broad \ion{He}{i}$\,\lambda\lambda5876,6678,7065$ emission, strong  emission in \Ha{}, very prominent emission between $\sim 5100$\,\AA{} and $\sim 5700$\,\AA{}, usually attributed to \ion{Fe}{ii} emission, as well as coronal line emission of [\ion{Fe}{vii}]$\,\lambda\lambda$5721,6087 and [\ion{Fe}{x}]$\,\lambda$6375. Due to the larger wavelength coverage, Spectrum 3 also reveals strong emission of the Balmer lines \Hg{} and \Hb{} as well as of \ion{He}{ii}$\,\lambda4686$ and the \ion{Fe}{ii} multiplet transitions 38 and 39 ($\sim 4500 - 4650$\,\AA{}).

Spectra 5 and 6 were obtained $95$ and $434$\,days after the transient peak, respectively, and reveal the fading of the strong blue continuum as well as of the broad emission lines. One notable exception from the general fading are the coronal lines [\ion{Fe}{vii}]$\,\lambda\lambda$5721,6087, which are stronger in the spectrum from 2018 October 4 than in the high-state spectra obtained $77$ and $67$\,days earlier. The spectrum from 2019 September 9 is approximately on the same level as the low-state spectrum from 2015 September 24, but still shows a slightly stronger continuum blueward of $\sim 6000$\,\AA{}.

\begin{figure*}[!h]
    \centering
    \includegraphics[width=1\textwidth,angle=0]{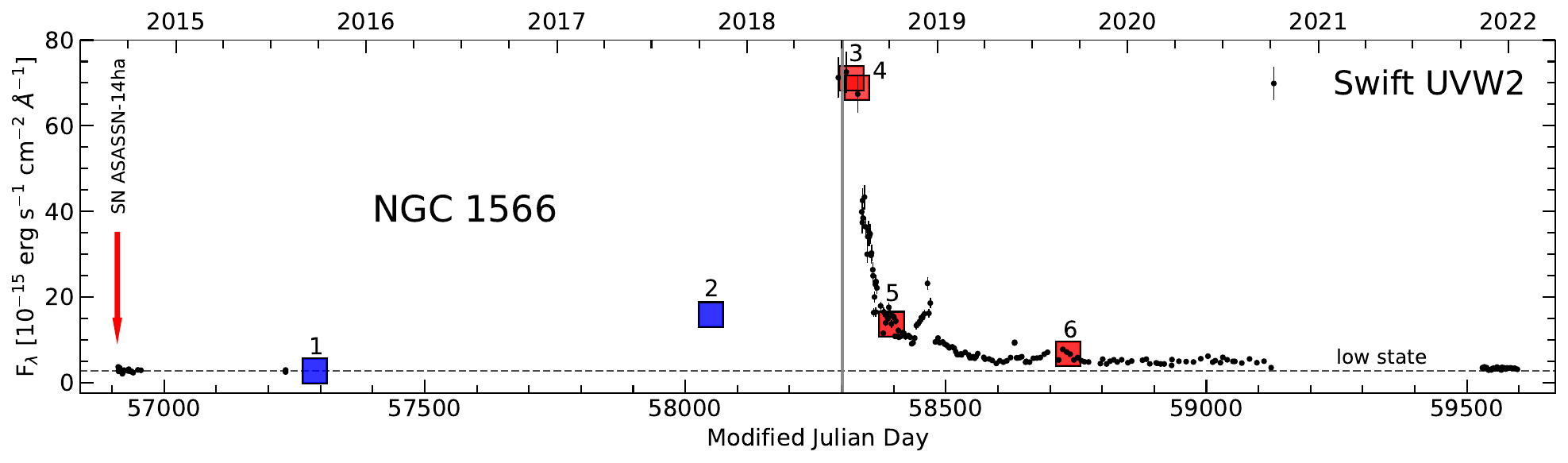}
    \caption{ Long-term UV \swift{} UVW2-band light curve before, during, and after the transient event in \ngc{} from 2017 to 2019. The blue and red boxes mark the time stamps of the spectroscopic MUSE and SALT observations, respectively, and are numbered chronologically. To guide the eye, the boxes are positioned such that they overlap with the UVW2 light curve; that is to say, they do not represent the actual optical flux values, but give a basic representation of the relative flux with respect to each other. The date of detection of the supernova ASASSN-14ha is indicated by a red arrow and the date of peak flux in the ASAS-SN light curve is shown by a gray line. The pretransient low-state flux level is indicated by a dashed black line.}
    \label{fig:Swift_W2_spectral}
\end{figure*}

To illustrate the timing of the spectral observations, we show the time stamps of all spectral observations along with the ultraviolet \swift{} UVW2-band light curve in Fig.~\ref{fig:Swift_W2_spectral}. Further, more detailed results on the spectral variations in \ngc{} during its transient event from 2017 to 2019 will be presented in future publications (Kollatschny et al., in prep.; Ochmann et al., in prep.).
%

%
\subsubsection{Host galaxy contribution }\label{sec:host_galaxy_contribution_results}
%
All spectra of \ngc{} (see Fig.~\ref{fig:all_spectra}) show a strong stellar signature from the underlying host galaxy. This holds especially true for the low-state spectra before and after the outburst, where the stellar signature of the host galaxy clearly dominates the continuum regions of the spectra. In order to determine the host galaxy contribution, we perform a spectral synthesis on Spectrum 1 from 2015 September 24. Of all the spectra in the campaign, this spectrum is the most suitable as it has the largest spectral coverage from $\sim 4700$\,\AA{} to 9300\,\AA{} and the lowest contribution from broad-line emission and nonstellar continuum.  The way we proceed is identical to that presented for IRAS\,23226-3843 by \citet{kollatschny23}. We use the Penalized Pixel-Fitting method (pPXF) \citep{cappellari04, cappellari17} and restrict the synthesis to wavelength ranges free from emission lines. This excludes in particular the \ion{Fe}{ii} complex at $\sim 5300$\,\AA{} from the fitting procedure. We used the stellar templates from the Indo-US library \citep{valdes04,shetty15,guerou17}, which provides high-enough spectral resolution, and fully covers the wavelength range of interest. In addition to the stellar templates, we add a constant component $F_{\lambda} =c$ mimicking a very weak power-law component $F_{\lambda} \propto \lambda^{-\beta}$ as the underlying nonstellar AGN continuum. This seems to us to be a reasonable estimate, since we cannot make an a priori statement about the nonstellar spectral index in the low-state spectrum and the contribution of a very weak power law can be approximated as constant in the optical regime.\footnote{We also note that when presented the possibility to choose from a range of power laws with different indices, pPXF always prefers the power law with the smallest index.}

\begin{figure*}[h!]
    \centering
    \includegraphics[width=1\textwidth,angle=0]{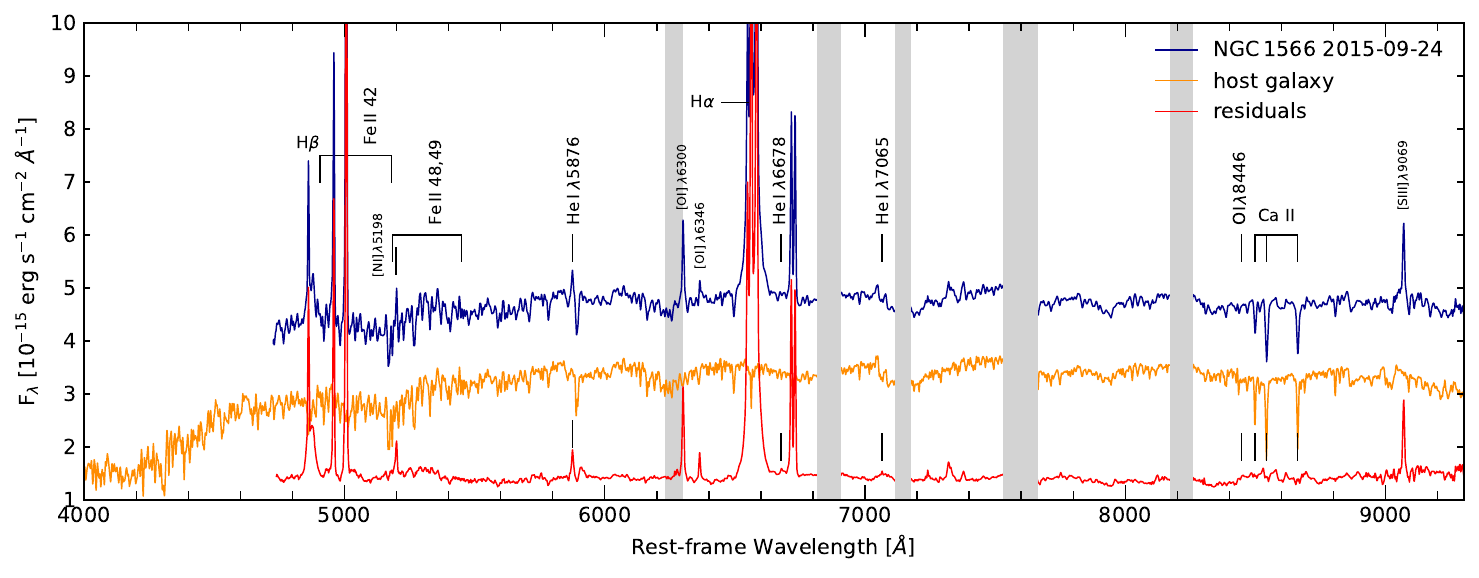}
    \caption{MUSE spectrum of \ngc{} taken on 2015 September 24 (Spectrum 1; blue) and the synthesis fit of the stellar contribution with pPXF (orange). The residuals (red) give the clean nuclear emission lines spectrum. For the fit, we flagged all prominent line emission including the \ion{Fe}{II} complex at $\sim 5300$\,\AA. The most prominent telluric absorption bands are flagged (gray).}
    \label{fig:spectral_synthesis}
\end{figure*}

The result of the spectral synthesis is shown in Fig.~\ref{fig:spectral_synthesis} together with the input spectrum and the residual spectrum, which is the clean nuclear spectrum of \ngc{} during its low state. The residual spectrum already includes the approximately constant nonstellar AGN component $F_{\lambda} = 1.4 \times 10 ^{-15}$ ergs s$^{-1}$ cm$^{-2}$\AA{}$^{-1}$. This allows us to estimate the host galaxy contribution in the original low-state spectrum to be  $\sim 60\%$ and $\sim 70\%$ in the B and V band, respectively. The clean low-state nuclear spectrum reveals line features formerly suppressed by the signature of the stellar population in the original spectrum. These features include, amongst others, narrow-line emission of [\ion{N}{i}]\,$\lambda$5198, \ion{Fe}{ii} emission of the transitions 42, 48, and 49, weak \ion{He}{i}\,$\lambda\lambda$5876, 6678, 7065 emission, and \ion{Ca}{ii}$\,\lambda\lambda$8498, 8542, 8662 triplet emission, as well as emission of \ion{O}{i}$\,\lambda$8446. 

pPXF determines a stellar velocity dispersion of $\sigma_\ast = 98^{+12}_{-9}$\,\kms{}. The exact value depends on the choice of the boundary conditions, namely the inclusion or exclusion of the NIR \ion{Ca}{ii} triplet and the probed wavelength region. We determined the error margins by reasonably varying the boundary conditions, that is, slighty varying the probed wavelength regions ($\pm 50$\,\AA{}), probing only the optical or NIR part, and exluding or including the NIR \ion{Ca}{ii} triplet, thereby obtaining a robust range of variation for $\sigma_\ast$. We note that the MUSE spectrum from 2017 October 23, although obtained under favorable seeing conditions, is not suitable to determine an estimation $\sigma_\ast$. In this spectrum, the most prominent absorption feature, namely the \ion{Ca}{ii} absorption triplet, is blended with \ion{Ca}{ii} emission, and many wavelength bands are affected by newly emerging line emission (see \ref{sec:optical_spectral_variations_results} and \ref{sec:NIR_spectral observations_results}). This limits the spectral range with a clean host-galaxy signature and introduces a large scatter in the distribution of determined stellar velocity dispersions $\sigma_\ast$.

%
\subsubsection{Balmer line profiles and their evolution}\label{sec:optical_profiles_results}
%

\begin{figure}[h!]
    \centering
    \includegraphics[width=0.47\textwidth,angle=0]{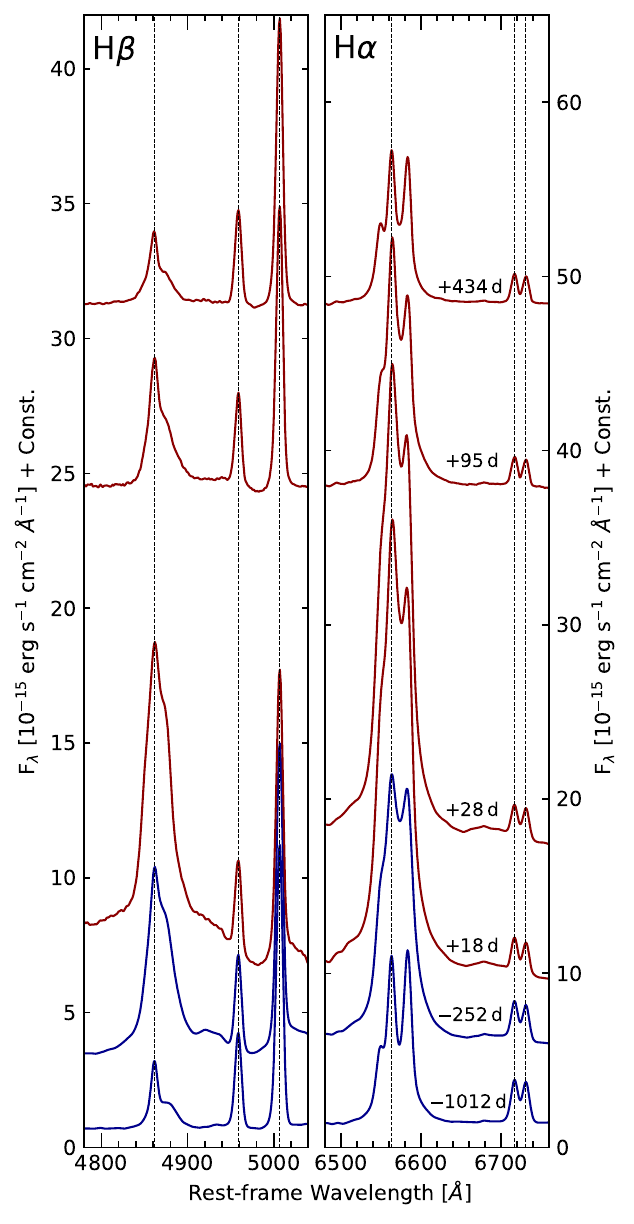}
    \caption{Temporal evolution (from bottom to top) of the host-free line profiles of \Hb{} (\textit{left panel}) and \Ha{} (\textit{right panel}). MUSE spectra are shown in blue, SALT spectra are shown in red. The profiles are shifted in flux for clarity. We indicate the central wavelengths of \Hb{} and \Ha{} by dashed lines. Likewise, we indicate the central wavelengths of the narrow lines [\ion{O}{iii}]$\,\lambda4959,5007$ and [\ion{S}{ii}]$\,\lambda\lambda6716,6731$ to demonstrate the accuracy of the spectral calibration.}
    \label{fig:Ha_Hb_wo_host}
\end{figure}

In order to obtain clean nuclear line profiles, we subtract the synthetic host-galaxy spectrum (see \ref{sec:host_galaxy_contribution_results}) from each spectrum after correcting all spectra to the same dispersion. The resulting host-free, singular-epoch line profiles of \Hb{} and \Ha{} are shown in Fig.~\ref{fig:Ha_Hb_wo_host}, where we indicate the central wavelength of \Hb{} and \Ha{} with a dashed line. In order to show the accuracy of the wavelength calibration, which is on the order of $\pm 20$\,\kms{}, we likewise indicate the central wavelengths of [\ion{O}{iii}]$\,\lambda4959,5007$ and [\ion{S}{ii}]$\,\lambda\lambda6716,6731$, respectively. Both the \Hb{} and the \Ha{} profiles show a pronounced redward asymmetry during all phases of the transient event, with no major changes in the overall line profile. This was also observed by \citet{alloin85}, who found the same redward asymmetry and no significant line profile variations despite considerable flux changes during their optical variability campaign of \ngc{} from 1980 to 1982. \citet{kriss91} reported redshifts of $200 - 1000\,$\kms{} for all broad lines in their UV to optical FOS/HST spectra.

\begin{figure}[!h]
    \centering
    \includegraphics[width=0.47\textwidth,angle=0]{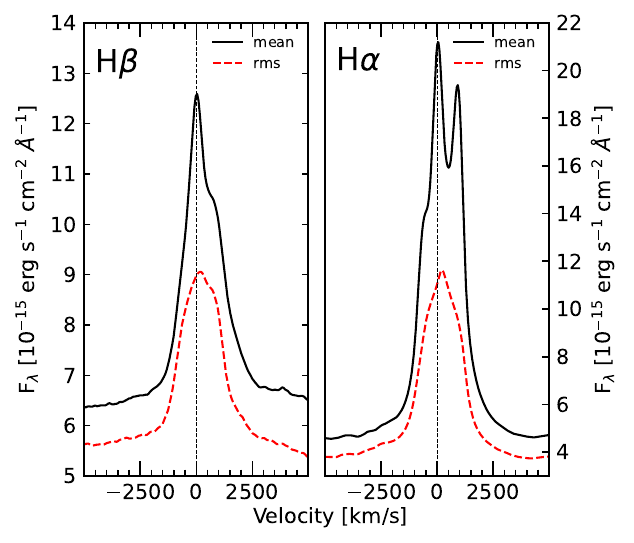}
    \caption{Mean (solid black) and rms (dashed red) line profiles of \Hb{} (\textit{left panel}) and \Ha{} (\textit{right panel}). The central velocity $v = 0$\,\kms{} is indicated by a black dashed line. The \Hb{} and \Ha{} rms profiles show a peak at $+(220 \pm 50)$\,\kms{} and $+(210 \pm 50)$\,\kms{}, respectively, with respect to the central wavelength. The profiles are strongly asymmetric with the red wing being broader by about $\sim 400$\,\kms{}. The FWHM amounts to $(2180 \pm 50)$\,\kms{} and $(2060 \pm 50)$\,\kms{} for \Hb{} and \Ha{}, respectively.}
    \label{fig:avg_rms_lines_wo_host}
\end{figure}

In order to assess the line profile variations of \Hb{} and \Ha{} in more detail, we calculate the host-free mean and rms line profiles. The resulting profiles are shown in Fig.~\ref{fig:avg_rms_lines_wo_host}. The \Hb{} and \Ha{} rms profiles, which map only the variable part of the line emission, show no evidence of residual narrow-line flux from  \Hb{}$_{\rm narrow}$, [\ion{O}{iii}]$\,\lambda\lambda4959,5007$ (see also \ref{sec:reconstructed_Balmer_profiles_results}) and \Ha{}$_{\rm narrow}$, [\ion{N}{ii}]$\,\lambda\lambda6548,6583$, respectively. This illustrates the high accuracy of the spectral intercalibration. The profiles have a FWHM of $(2180 \pm 50)$\,\kms{} and $(2060 \pm 50)$\,\kms{} for \Hb{} and \Ha{}, respectively, and are strongly asymmetric with respect to the rest-frame velocity, with the red wing being broader by about $\sim 400$\,\kms{} with respect to the central wavelength. In addition, both rms profiles show an additional narrow peak component that is not associated with narrow-line residuals, but instead is shifted with respect to the rest-frame central wavelength by $(220 \pm 50)$\,\kms{} and $(210 \pm 50)$\,\kms{}, respectively. With respect to the peak positions, the central rms profiles are almost perfectly symmetric. Major deviations from symmetry are only evident in the extended line wings.

At this point, the individual \Hb{} and \Ha{} line profiles still comprise contributions from the narrow components \Hb{}$_{\rm narrow}$ and \Ha{}$_{\rm narrow}$ as well as [\ion{N}{ii}]\,$\lambda\lambda$6548, 6583, respectively. Therefore, in order to obtain clean FWHM measurements for singular epochs, we subtract a scaled [\ion{O}{iii}]\,$\lambda 5007$ profile taken from the 2015 September 24 spectrum as a mean template for each narrow-line component from the total line profile. We adopt this procedure as we explicitly assume that the narrow-line components are not purely Gaussian, but instead are more complex as they are being shaped by the kinematics of the narrow-line region. This is supported by the findings of \citet{alloin85} and \citet{silva18}, who found that adequately modeling the narrow lines in \ngc{} requires at least two Gaussians. 

\begin{figure}[h!]
    \centering
    \includegraphics[width=0.47\textwidth,angle=0]{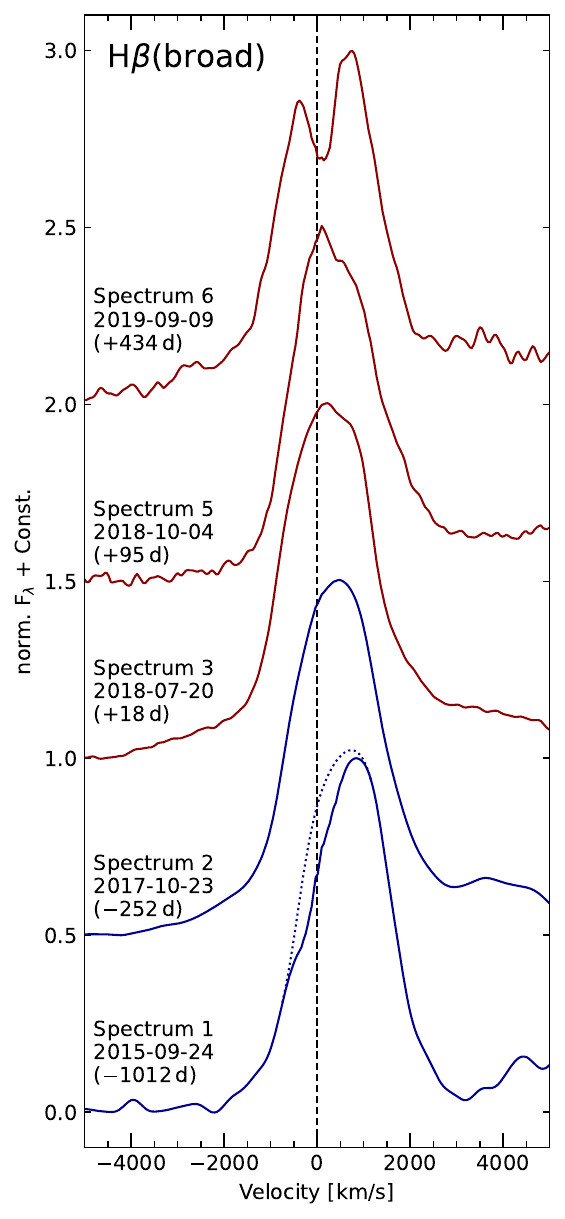}
    \caption{Temporal evolution (from bottom to top) of the normalized, narrow-line, and host-galaxy-subtracted \Hb{} profiles in velocity space. A suitable linear pseudo-continuum was subtracted, and the spectra are flux-shifted for clarity. MUSE spectra are shown in blue, SALT spectra are shown in red. The central velocity $v = 0$\,\kms{} is indicated by a dashed line. We give the spectrum ID, the date of observations, and the time in days with respect to the peak time $t_0 = 58301.44$\,MJD of the transient event. The reconstructed \Hb{} profile for  Spectrum 1 (2015 September 24) is shown as a dotted line (see \ref{sec:reconstructed_Balmer_profiles_results}).}
    \label{fig:clean_broad_Balmer_profiles}
\end{figure}

\begin{figure}[h!]
    \centering
    \includegraphics[width=0.47\textwidth,angle=0]{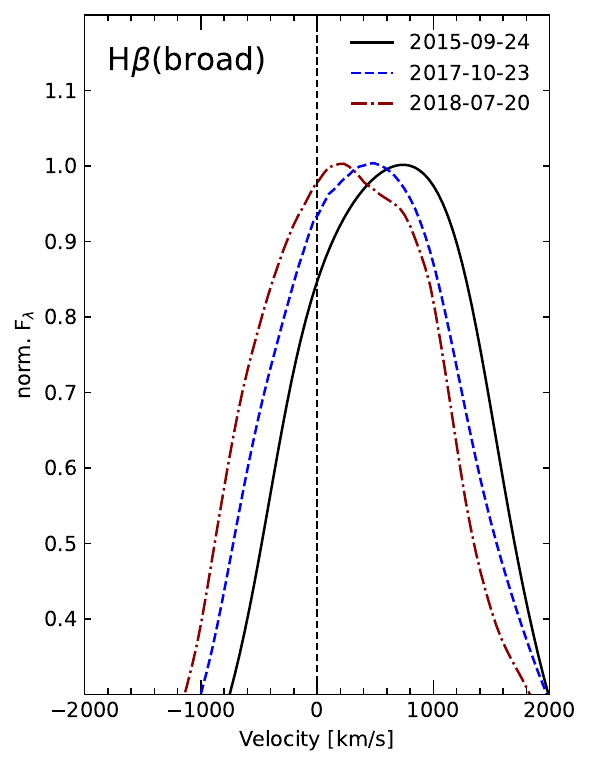}
    \caption{Blueward drift of the normalized \Hb{} profiles in velocity space (from Spectrum 1 (2015 September 24) to Spectrum 2 (2017 October 23), and Spectrum 3 (2018 July 20)) after subtraction of the narrow-line and the host-galaxy contribution. The \Hb{} profile from Spectrum 1 is the reconstructed \Hb{} profile (see \ref{sec:reconstructed_Balmer_profiles_results}). The blueward drift of the \Hb{} profile during the rising phase is clearly visible. The redshift of the \Hb{} line shifts from $+(730 \pm 50)$\,\kms{} to $+(490 \pm 50)$\,\kms{} and $+(360 \pm 50)$\,\kms{} from Spectrum 1 to Spectrum 2 and Spectrum 3, respectively.}
    \label{fig:Hbeta_blueward_drift}
\end{figure}

We show the resulting broad \Hb{} line profiles after subtraction of a suitable linear pseudo-continuum in Fig.~\ref{fig:clean_broad_Balmer_profiles}. From each of these profiles, we subtracted a constant narrow-line component \Hb{}$_{\rm narrow}$ with a flux of $18.2 \, \times$\,10$^{-15}$ ergs s$^{-1}$ cm$^{-2}$. We give the measured FWHM and redshift of all \Hb{}$_{\rm broad}$ profiles in Table~\ref{tab:FWHM}. We observe the following trends in the emission lines: The \Hb{} profiles in Spectrum 1, Spectrum 2 and Spectrum 3 are skewed and clearly display the redward asymmetry mentioned previously. While the \Hb{}$_{\rm broad}$ profile in Spectrum 1 exhibits minor distortions of the central profile, probably due \Hb{}$_{\rm narrow}$ residuals (see also \ref{sec:reconstructed_Balmer_profiles_results}), the profile in Spectrum 2 appears to be free from this effect. Most strikingly, we observe a substantial change in redshift of the \Hb{}$_{\rm broad}$ profile between Spectrum 1 and Spectrum 2, with a shift of the line peak from $+(730 \pm 50)$\,\kms{} to $+(490 \pm 50)$\,\kms{}. This trend continues until Spectrum 3, where the redshift of the profile only amounts to $+(360 \pm 50)$\,\kms{}.  We term this velocity shift of the total \Hb{} profile during the rising phase of the transient event to be a \textit{blueward drift} of the line profile. To illustrate we show the normalized \Hb{} profiles from Spectrum 1 to Spectrum 3 in Fig.~\ref{fig:Hbeta_blueward_drift}.

In comparison to the \Hb{} profiles from Spectrum 1 to Spectrum 3, the \Hb{} profile in Spectrum 5 is slightly distorted with an apparent additional emission component at $+(130 \pm 10)$\,\kms{}. The redshift increases from $+(360 \pm 50)$\,\kms{} to $+(450 \pm 100)$\,\kms{} from Spectrum 3 to Spectrum 5. The profile in Spectrum 6 is two-peaked, caused by the apparent emission component in Spectrum 5 now being present as an  apparent absorption component that distorts the profile. Due to the absorption, the \Hb{}$_{\rm broad}$ profile cannot be normalized to peak height like the previous profiles, and no meaningful measurement of the redshift can be performed.

\begin{table}[h!]
\caption{FWHM and redshift $\Delta v$ (with respect to the central wavelength) of the \Hb{} line profile for all epochs.}
\centering
\resizebox{0.48\textwidth}{!}{
   \begin{tabular}{llcrlc}
        \hline \hline
        \noalign{\smallskip}
        ID & \mcc{MJD} & UT Date & $(t-t_0)$ & \mcc{FWHM$_{H\beta}$}     & $\Delta v_{H\beta}$    \\
           &         &         & \mcc{[days]}  &\mcc{[\kms{}]}   &  \mcc{[\kms{}]}        \\
        \noalign{\smallskip}
        \hline 
        \noalign{\smallskip}
        1 & 57289.23       &   2015-09-24  & $-1012$   & $1970 \pm 50$     &   $+730 \pm 50$  \\
        2 & 58049.20       &   2017-10-23  & $-252$    & $2270 \pm 100$    &   $+490 \pm 50$  \\
        3 & 58319.17            &       2018-07-20      & $+18$     & $2230 \pm 100$        &   $+360 \pm 50$  \\
        4 & 58329.16       &   2018-07-30  & $+28$     & \mcc{- }          &   -              \\
        5 & 58395.96            &       2018-10-04      & $+95$     & $2100 \pm 100$    &   $+450 \pm 100$  \\
        6 & 58735.03            &       2019-09-09      & $+434$    & $2440 \pm 300$        &       -               \\

        \hline
    \end{tabular}}
\tablefoot{Spectrum 4 from 2018 July 30 does not cover the \Hb{} line. The \Hb{} profile in Spectrum 6 from 2019 September 09 is distorted and no meaningful normalization of the profile is possible with respect to the other profiles.
}    
\label{tab:FWHM}
\end{table}

Strikingly, the profile and the FWHM of \Hb{}$_{\rm broad}$ do not change significantly during the rising phase of the transient event. The slightly lower value of FWHM $=(1970 \pm 50)$\,\kms{} in the low-state Spectrum 1 from 2015 September 24 compared to the other spectra is caused by the minor distortion of the central profile due to a narrow-line residual. Taking this residual into account (see \ref{sec:reconstructed_Balmer_profiles_results}), the width amounts to FWHM $= (2200 \pm 50)$\,\kms{}, and is therefore in perfect agreement with the values obtained for the other profiles.\footnote{We also use the reconstructed profile from \ref{sec:reconstructed_Balmer_profiles_results} to determine the redshift of the line.} 

Although our procedure is able to recover clean \Hb{}$_{\rm broad}$ profiles, it is not successful in recovering \Ha{}$_{\rm broad}$ profiles. This is due to differences in the exact profile shape and width between [\ion{O}{iii}]$\,\lambda5007$ and [\ion{N}{ii}]\,$\lambda\lambda$6548, 6583. Nevertheless, because of the very similar rms profiles of both \Hb{} and \Ha{}, we suspect \Ha{}$_{\rm broad}$ to show the same behavior as \Hb{}$_{\rm broad}$.

%
\subsubsection{Black hole mass estimation using the M$_{\rm BH}-\sigma_{\ast}$ and M$_{\rm BH}$ $-$  FWHM(\Hb{}), L$_{\rm 5100}$ scaling relations} \label{sec:black_hole_mass_results}
%
In \ref{sec:host_galaxy_contribution_results}, we obtain a value of $\sigma_\ast = 98^{+12}_{-9}$\,\kms{} for the stellar velocity dispersion in the nuclear region of \ngc{} during its low state. For the same spectrum we obtain a clean measurement of FWHM(\Hb{})$=(2200 \pm 50)$\,\kms{} in \ref{sec:optical_profiles_results}. These results allow us to estimate the black hole mass M$_{\rm BH}$ using the M$_{\rm BH}-\sigma_{\ast}$ scaling relation of \citet{onken04} (see their Eq.\ 2) and the M$_{\rm BH}$ $-$  FWHM(\Hb{}), L$_{\rm 5100}$ scaling relation of \citet{vestergaard06} (see their Eq.\ 5). From the M$_{\rm BH}-\sigma_{\ast}$ scaling relation we obtain
\begin{equation}
    M_{\rm BH,\, \sigma_{\ast}} = 4.4^{+6.7}_{-2.3} \times 10^6\,M_{\odot}.
\end{equation}
For the M$_{\rm BH}$ $-$  FWHM(\Hb{}), L$_{\rm 5100}$ scaling relation, we measure L$_{\rm 5100}$ in the host-free low-state spectrum. We obtain a continuum flux of F$_{\lambda} = 1.4 \times $10$^{-15}$\,ergs s$^{-1}$ cm$^{-2}$ \AA{}$^{-1}$, which results in a luminosity of $\lambda$L$_{\rm 5100} = 3.91 \times 10^{41}$\, ergs s$^{-1}$. Together with FWHM(\Hb{}) $=(2200 \pm 50)$\,\kms{} measured in \ref{sec:optical_profiles_results}, we therefore obtain a black hole mass of 
\begin{equation}
    M_{\rm BH,\, FWHM(H\beta),\, L_{\rm 5100}} = 2.5^{+0.2}_{-0.3} \times 10^6\,M_{\odot}.
\end{equation}

Using a velocity dispersion of $\sigma_{\ast} = (105 \pm 10)\,$\kms{}, which is slightly higher than $\sigma_\ast = 98^{+12}_{-9}$\,\kms{} obtained by us, \citet{smajic15} obtained a mass of $M_{\rm BH} = (8.6 \pm 4.4) \times 10^6\,M_{\odot}$.  Using also the flux and FWHM of broad Br$\,\gamma$ from their data, they estimated the black hole mass $M_{\rm BH}$ in \ngc{} to be $M_{\rm BH} = (3.0 \pm 0.9) \times 10^6\,M_{\odot}$, and found their results to be in good agreement with results obtained by \citet{woo02}  and \citet{kriss91}, respectively. We give their results and our values in Table~\ref{tab:M_SMBH}. From here on, we adopt the mean black hole mass of $M_{\rm SMBH} = (5.3 \pm 2.7) \times 10^6$\,M$_{\odot}$ for \ngc{}.

\begin{table}[h!]
\caption{Black hole masses $M_{\rm SMBH}$ for \ngc{} determined by different studies. For clarity, we denote if the stellar velocity dispersion $\sigma_{\ast}$ or the FWHM was used for mass determination.}
\centering
    \begin{tabular}{lcr}
        \hline \hline
        \noalign{\smallskip}
           \mcc{$M_{\rm SMBH}$}              & Method  &   \mcc{Reference}  \\ \
           [$10^{6}$\,M$_{\odot}$]      &   &               \\
        \noalign{\smallskip}
        \hline 
        \noalign{\smallskip}
           $4.4\,\,\,\,^{+6.7}_{-2.3}$         & $\sigma_{\ast}$   &   This work\\ [0.5em]
           $2.5\,\,\,\,^{+0.2}_{-0.3}$         & FWHM       &   This work\\ [0.5em]
           $8.6 \pm 4.4$                       & $\sigma_{\ast}$   &   \citet{smajic15}\\ [0.5em]
           $3.0 \pm 0.9$                       & FWHM       &   \citet{smajic15}\\ [0.5em]
           $8.3$                               & $\sigma_{\ast}$   &  \citet{woo02}\\ [0.5em]
           $5$                                 & FWHM       &  \citet{kriss91}\\
        \hline
        \mcc{mean} & & \\
         $5.3 \pm 2.7$ & &\\
        \hline
    \end{tabular}
\label{tab:M_SMBH}
\end{table}
%

%
\subsection{Near-infrared spectral observations}\label{sec:NIR_spectral observations_results}
\begin{figure*}[!h]
    \centering
    \includegraphics[width=1\textwidth,angle=0]{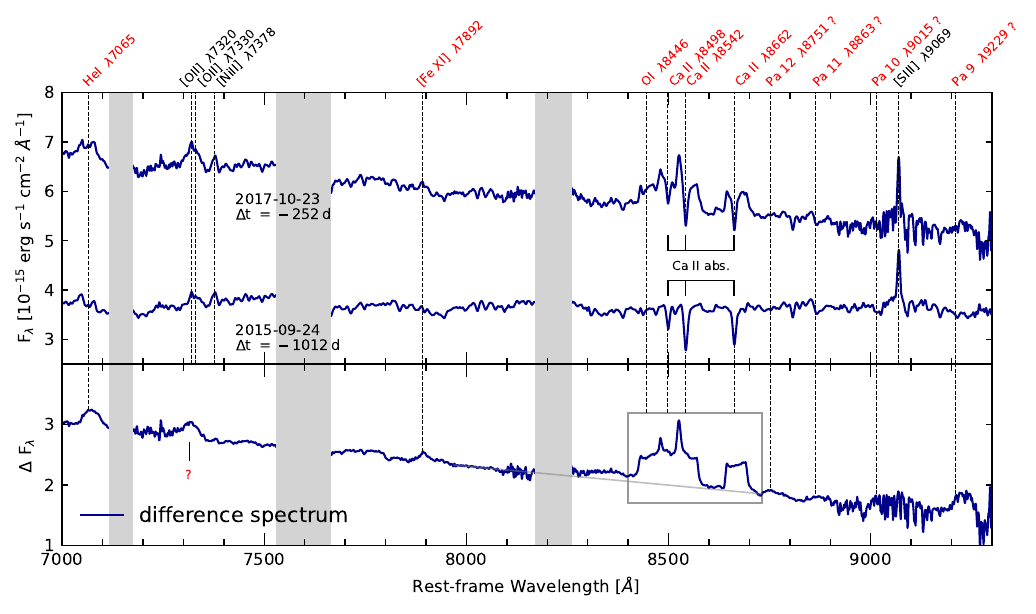}
    \caption{Intercalibrated MUSE NIR spectra from 2015 September 24 and 2017 October 23 (\textit{upper panel}) and the resulting difference spectrum (\textit{lower panel}). The most prominent telluric absorption bands are shown in gray. In addition to an increase in NIR continuum flux, the difference spectrum reveals emission from several broad lines and [\ion{Fe}{xi}]$\,\lambda 7892$. The most prominent emission feature is the blend of \ion{O}{i}$\,\lambda$8446 and the \ion{Ca}{ii}$\,\lambda\lambda$8498, 8542, 8662 triplet (gray box). The linear pseudo-continuum used for later analysis is shown as a gray line. The positions of all identified emission lines are marked by dashed lines. Narrow emission lines and the stellar \ion{Ca}{ii} triplet absorption are denoted in black, while variable line emission is denoted in red. }
    \label{fig:MUSE_NIR_spectra}
\end{figure*}

%
MUSE observed \ngc{} on 2015 September 24 and 2017 October 23 in the wavelength range $\sim 4700$\,\AA\ to $9300$\,\AA{}. This is $\sim 700$\,days before and $\sim 50$\,days after the start of the reported brightening of \ngc{} in September 2017, respectively. We show the NIR part (rest-frame wavelength $7000$\,\AA{} -- $9300$\,\AA{}) of the nuclear spectra together with the resulting difference spectrum in Fig.~\ref{fig:MUSE_NIR_spectra}. The spectra were extracted using a circular aperture with a radius of 1\arcsec{}. In contrast to the optical regime, the spectra are intercalibrated to the same narrow-line flux of [\ion{O}{ii}]$\,\lambda\lambda7320,7330$, [\ion{Ni}{ii}]$\,\lambda7378$, and [\ion{S}{iii}]$\,\lambda9069$, as well as the same absorption strength in the \ion{Ca}{ii}$\,\lambda\lambda$8498, 8542, 8662 triplet. The absorption strength of the \ion{Ca}{ii} triplet can be considered constant due to the identical aperture of both observations, in other words, the underlying stellar population from the host galaxy is identical.

Spectrum 1 from 2015 September 24 is clearly dominated by the stellar contribution from the host galaxy. The most prominent emission feature in this spectrum are the narrow lines [\ion{O}{ii}]$\,\lambda\lambda7320,7330$, [\ion{Ni}{ii}]$\,\lambda7378$ and [\ion{S}{iii}]$\,\lambda9069$. In addition, prominent absorption in the \ion{Ca}{ii}$\,\lambda\lambda$8498, 8542, 8662 triplet is present. Spectrum 2 from 2017 October 23, approximately 50\,days into the brightening, shows the emergence of a nearly linear continuum across the entire wavelength band, as well as several additional emission lines. Both continuum and line flux can be best recognized in the difference spectrum. We identify broad emission from \ion{He}{i}$\,\lambda7065$, possible emission from Pa\,12$\,\lambda8751$, Pa\,11$\,\lambda8863$ (although only marginally detected in both cases) and Pa\,9$\,\lambda9229$, as well as emission from the coronal line [\ion{Fe}{xi}]$\,\lambda7892$, all of which previously not present in the low-state spectrum from 2015 September 24. In addition to the lines identified before, we observe the emergence of broad emission at $\sim 7306$\,\AA{}, which cannot be unambiguously identified. This broad-line feature might also be present in other NIR AGN spectra \citep[e.g.,][]{landt08}, however, a clear detection in singular-epoch spectra is difficult due to blending with the narrow lines [\ion{O}{ii}]$\,\lambda\lambda7320,7330$.

The most prominent emission feature is broad emission in \ion{O}{i}$\,\lambda8446$ and the \ion{Ca}{ii}$\,\lambda\lambda$8498, 8542, 8662 triplet. We indicate these lines in Fig.~\ref{fig:MUSE_NIR_spectra} by a gray box. The linear continuum subtracted from the line profiles for further analysis is shown by a gray line. A thorough analysis of the line profiles is performed in \ref{sec:CaII_profile_results},  \ref{sec:fitting_results} and \ref{sec:OI_CaII_complex_profile_results}. We show that the \ion{Ca}{ii}$\,\lambda8662$ profile is well approximated by emission from an elliptical disk. Furthermore, the blended total profile of \ion{O}{i}$\,\lambda$8446 and \ion{Ca}{ii}$\,\lambda\lambda$8498, 8542 can be reconstructed using the \ion{Ca}{ii}$\,\lambda8662$ difference profile as a template for all three lines.

%
\subsubsection{The double-peaked \ion{Ca}{ii}\,$\lambda 8662$ line profile}\label{sec:CaII_profile_results}
%

We show the clean line profile of \ion{Ca}{ii}$\,\lambda8662$ in velocity space (after subtraction of the underlying linear continuum indicated in Fig.~\ref{fig:MUSE_NIR_spectra}) in Fig.~\ref{fig:CaII_8662_veloplot}. The \ion{Ca}{ii}$\,\lambda8662$ difference profile is double-peaked, with the red peak being stronger than the blue peak by about 10\%. Moreover, the red peak is composed of three individual subpeaks that form a ``trident'' structure. The total profile is strongly asymmetric with the right wing being broader by $\sim 300$\,\kms{}. The full width at half maximum (FWHM) amounts to $ (1920 \pm 50)$\,\kms{} when the \ion{Ca}{ii}\,$\lambda 8662$ profile is normalized to ${\rm F}_\lambda(0$\,\kms{}$) = 1$. The blue and red peak are positioned at $v_{\rm blue peak} \approx -615$\,\kms{} and $v_{\rm red peak} \approx +950$\,\kms{}, respectively, and are therefore separated by $\sim 1600$\,\kms{}.

The profile closely resembles that of \ion{H}{i} 21\,cm-line emission profiles in so-called ``lopsided'' galaxies, where the matter distribution in the galaxy's plane is asymmetric with respect to the galaxy's center \citep[e.g.,][]{richter94}. Similar asymmetric line profiles have been observed in a number of AGN, for example, Arp\,102B \citep{popovic14}, NGC\,4958 \citep{ricci19}, NGC\,1097 \citep{storchi-bergmann97, schimoia15}, and others \citep{gezari07, lewis10}, as well as TDEs, for example, PTF09djl \citep{liu17}, AT\,2018hyz \citep{hung20}, AT\,2020zso \citep{wevers22}, and are thought to be signatures of an elliptical (accretion) disk. We analyze the line profile of \ion{Ca}{ii}\,$\lambda 8662$ in the framework of an elliptical accretion disk model in \ref{sec:fitting_results} in more detail.

%
\subsubsection{Fitting the \ion{Ca}{ii}\,$\lambda 8662$ difference profile with an elliptical accretion disk model}\label{sec:fitting_results}
%
The analysis of the \ion{Ca}{ii}\,$\lambda 8662$ difference profile reveals a strongly asymmetric, double-peaked profile with the red peak being stronger than the blue peak by about 10\%. Emission line profiles with stronger red than blue peaks are inconsistent with circular accretion disk models \citep[see, e.g.,][]{eracleous95, lewis10}. Instead, they require an asymmetric distribution of matter (or emissivity) in the accretion disk such that the receding part of the disk -- with respect to the observer -- contributes more to the line flux than the approaching part. We note, however, that observed double-peaked line profiles attributed to the emission from an accretion disk are generally more complex and often additional components such as Gaussians are needed to obtain good fits \citep[e.g.,][]{hung20, wevers22}. In general, line profiles in AGN are most likely shaped by a superposition of several effects, such as, amongst many others, the geometry of the BLR, turbulence, (disk) winds \citep[e.g.,][]{goad96, schulz95, goad12, flohic12} or, in the case of \ion{Ca}{ii}$\,\lambda\lambda8498, 8542,8662$ triplet emission, stellar absorption.

We assume that the \ion{Ca}{ii}$\,\lambda8662$ profile is indeed a genuine double-peaked profile ; that is to say, it is not caused by the underlying stellar absorption of the host galaxy (see \ref{sec:robustness_line_profile_discussion} for more details). We therefore fit the \ion{Ca}{ii}\,$\lambda 8662$ line profile with the relativistic elliptical accretion disk model of \citet{eracleous95}, in which the total observed flux $F$ from the line is described by the expression
\begin{equation}
    F = \int d\nu \int \int d\Omega\, I_{\nu}
,\end{equation}
where $\nu$, $d\Omega$, and $I_{\nu}$ are the frequency, solid angle element as seen by the observer, and the specific intensity. The specific intensity $I_{\nu_e}$ in the frame of the emitter is given as
\begin{equation}
    I_{\nu_e} = \frac{1}{4\pi}\frac{\epsilon_0\, \xi^{-q}}{\sqrt{2\pi}\sigma} \exp{\left[- \frac{(\nu_e - \nu_0)^2}{2\sigma^2} \right],}
\end{equation}
where $\sigma$ is the broadening parameter, $\epsilon(\xi)=\epsilon_0\, \xi^{-q}$ is the line emissivity, and $\nu_e$ and $\nu_0$ are the emitted and rest frequency, respectively. The model has seven free parameters, namely the inner and outer pericenter distance $\xi_1$ and $\xi_2$, the inclination angle $i$, the major axis orientation $\phi_0$, the broadening parameter $\sigma$, the disk eccentricity $e$, and the emissivity power-law index $q$ (see \citet{eracleous95} for details). 

To find the best-fitting parameter set to the data, we apply a combination of the Monte Carlo method and momentum-based gradient descent to minimize $\chi^2$ of the model fit. In a first step, we flag the central part of the profile between $-450\,$\kms{} $< v < +850\,$\kms{}, as well as the region for which $+1350\,$\kms{} $< v < +1750\,$\kms{}, that is, the inner two peaks of the trident and the outermost red line wing. We proceed in this way as we assume the red trident peak to be a superposition of three individual peaks, all of them with the same width as the blue peak of the \ion{Ca}{ii}$\,\lambda8662$ profile (see \ref{sec:disk_inhomogeneities_discussion} for details). We then restrict the posterior parameter space by creating $\sim 250\,000$ models by sampling from a uniform prior parameter distribution of $200\,r_g < \xi_1 < 10\,000\,r_g$, $500\, r_g < \xi_2 < 20\,000\,r_g$, $0^{\circ} < i < 20^{\circ}$, $0^{\circ} \leq \phi_0 < 360^{\circ}$, $0\,$\kms{} $< \sigma < 2000\,$\kms{}, $0 \leq e < 1$, and $ 1 < q < 5$, from which the parameters are randomly drawn.The choice for the inclination $i$ is based on the results of \citet{parker19}, who determined an inclination angle of $i<11^{\circ}$. To scan the parameter space, we first vary individual parameters while leaving the rest of the parameter set fixed. We start with the inclination angle $i$ and the broadening parameter $\sigma$, which mainly govern the width of the total profile and of the blue and red peak, and then investigate the effects of the major axis orientation $\phi_0$, the inner and outer pericenter distance $\xi_1$ and $\xi_2$, and finally the disk eccentricity $e$ and the emissivity power-law index $q$. We then examine the parameter space in more detail by iteratively increasing the number of free parameters (up to seven) and varying them with an increasingly finer parameter grid. This iterative approach is similar to the procedure presented by \citet{short20}.

\begin{figure}[h!]
    \centering
    \includegraphics[width=0.47\textwidth,angle=0]{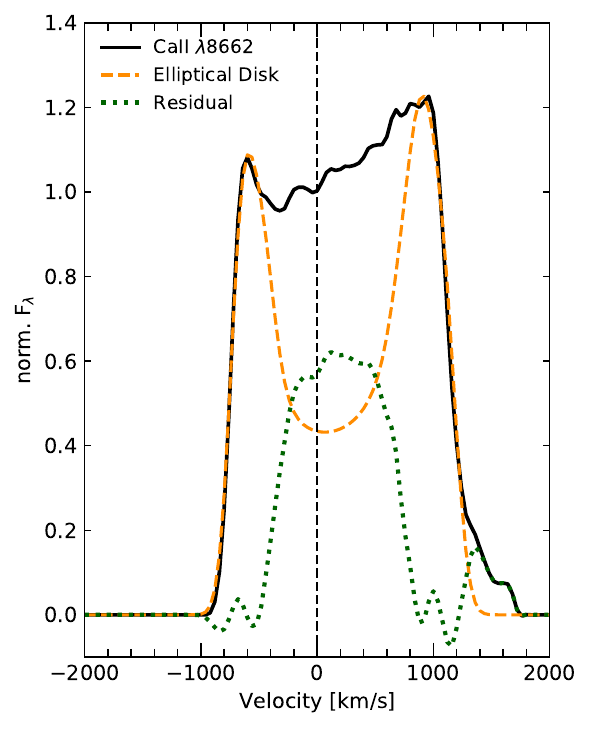}
    \caption{Normalized difference line profile of \ion{Ca}{ii}\,$\lambda 8662$ in velocity space (\textit{black}) after subtraction of a linear pseudo-continuum. The profile is double-peaked and strongly asymmetric, with the red peak being stronger (by about 10\%) and being shifted to higher velocities ($v_{\rm red peak} \approx +950$\,\kms{}) than the blue peak ($v_{\rm blue peak} \approx -615$\,\kms{}). In addition, the red peak shows a ``trident'' structure, meaning, it is composed of three individual subpeaks. The width of the profile amounts to $ {\rm FWHM} = (1920 \pm 50)$\,\kms{}. We show the best-fit results for the elliptical disk model (\textit{orange}) and the residual flux (\textit{green}).}
    \label{fig:CaII_8662_veloplot}
\end{figure}

We find that the only reasonable solutions reproducing the key features of the line profile (namely FWHM, peak width, and relative peak height) require $1000\,r_g < \xi_1 < 5000\,r_g$, $3000\, r_g < \xi_2 < 8000\,r_g$, $5^{\circ} < i < 12^{\circ}$, $150^{\circ} \leq \phi_0 < 250^{\circ}$, $50\,$\kms{} $< \sigma < 200\,$\kms{}, $0.1 \leq e < 0.8$, and $ 2 < q < 5$. In a second step, we apply the momentum-based gradient descent method to find the minimal value for $\chi^2$. In order to exclude running into local minima for $\chi^2$, we repeat the run with the momentum-based gradient descent method $1\,000$ times with start parameters randomly drawn from the restricted prior parameter distribution. We give the best-fit parameter set of the model to our data, in other words, the parameter set from the run that realizes the minimal $\chi^2$, in Table~\ref{tab:bestfit_parameters} and show the resulting model line profile in Fig.~\ref{fig:CaII_8662_veloplot}. The \ion{Ca}{ii}$\,\lambda8662$ profile is best modeled with an almost face-on, $i = (8.10 \pm 3.00)^{\circ}$, but eccentric accretion disk with an eccentricity of $e = (0.57 \pm 0.35)$, viewed under a major axis orientation of $\phi_0 = (193.29 \pm 26.00)^{\circ}$. The internal broadening to turbulence is low with $\sigma = (87 \pm 10)$\,\kms{} (corresponding to $v_{\rm turb} = 200$\,\kms{}), and the emissivity power-law index is $q = (4.34 \pm 0.80)$.
\begin{table*}[ht!]
\caption{Best-fit parameter set for the \ion{Ca}{ii}\,$\lambda 8662$ profile fit using the elliptical disk model of \citet{eracleous95}.}
\centering
    \begin{tabu} to \textwidth {@{} *7{X[c]} @{}}
    
        \hline \hline
        \noalign{\smallskip}
        $\xi_1$    &   $\xi_2$    &   $i$ &    $\phi_0$ &  $\sigma$    &   $e$ &   $q$ \\ \
        [r$_g$]    &   [r$_g$]    &   [deg] &   [deg]  &    [km/s]     &       &       \\ 
        \noalign{\smallskip}
        \hline 
        \noalign{\smallskip}
        $2231 \pm 1000$    &   $4050 \pm 1500$    &   $8.10 \pm 3.00$ &   $193.29 \pm 26.00$    &   $87 \pm 10$ &   $0.57 \pm 0.35$ &   $4.34 \pm 0.80$\\
        \hline 
    \end{tabu}
\label{tab:bestfit_parameters}
\end{table*}
The blue wing of the \ion{Ca}{ii}$\,\lambda8662$ profile is very well approximated by the elliptical disk model, minor deviations are only found at the base of the wing and in the exact height of the peak. While the red wing is in general also well approximated by the best-fit parameter model, the red peak of the model is shifted slightly inwards by about 50\,\kms{}. The central part of the profile between $-450\,$\kms{} $< v < +850\,$\kms{} is not well approximated by the purely elliptical disk model, and an additional component is clearly visible in the difference profile in Fig.~\ref{fig:CaII_8662_veloplot}. We do not fit the complete line profile by including an additional Gaussian component as has been done in other studies \citep[e.g.,][]{hung20, wevers22}, since this additional component is clearly not a Gaussian. A thorough discussion of the best-fit results is given in \ref{sec:disk_origin_discussion}.

%
\subsubsection{Decomposing the blended \ion{O}{i} and \ion{Ca}{ii} profile}\label{sec:OI_CaII_complex_profile_results}
\begin{figure}[h!]
    \centering
    \includegraphics[width=0.47\textwidth,angle=0]{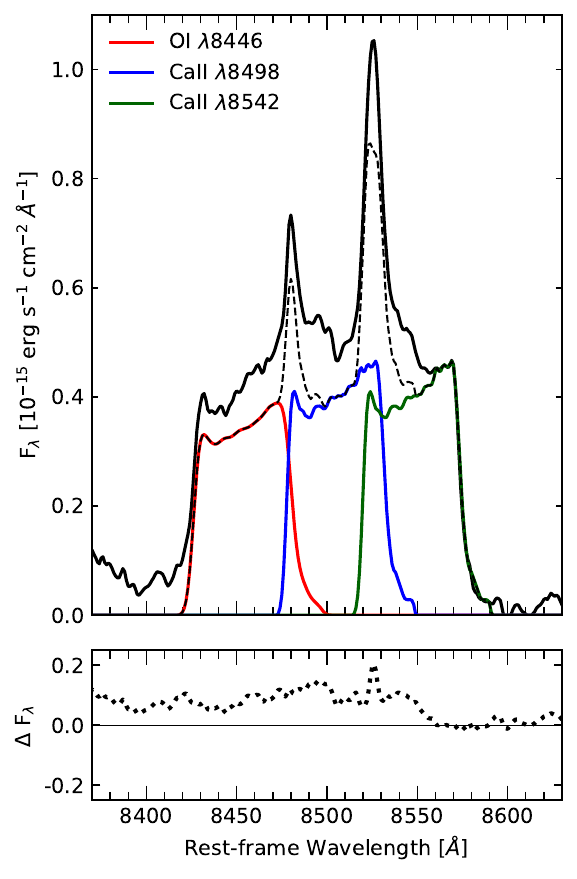}
    \caption{Decomposition of the \ion{O}{i}$\,\lambda$8446 and \ion{Ca}{ii}$\,\lambda\lambda$8498, 8542 complex using the \ion{Ca}{ii}$\,\lambda 8662$ line profile (see Fig.~\ref{fig:CaII_8662_veloplot}) as a template for each line constituting the blended profile (black solid line; \textit{upper panel}). We are able to reconstruct the \ion{O}{i}$\,\lambda$8446 and \ion{Ca}{ii}$\,\lambda\lambda$8498, 8542 complex using a \ion{Ca}{ii} triplet ratio of 1:1:1 and an \ion{O}{i}$\,\lambda$8446-to-\ion{Ca}{ii}\,$\lambda 8662$ ratio of 0.85:1. In order to be able to cleanly reconstruct the slope of the blue wing of \ion{O}{i}$\,\lambda$8446, we have to convolve the \ion{Ca}{ii}\,$\lambda 8662$ template with a Gaussian with a width corresponding to $\sigma = 70$\,\kms{}. The difference between original blended line profile (black solid line) and reconstructed line profile (black dashed line) is shown in the \textit{lower panel}.}
    \label{fig:OI_CaII_complex}
\end{figure}

%
We show the blended line profile of \ion{O}{i}$\,\lambda8446$ and the \ion{Ca}{ii}$\,\lambda\lambda$8498, 8542 lines (after subtraction of the same linear continuum as for \ion{Ca}{ii}$\,\lambda$8662) in Fig.~\ref{fig:OI_CaII_complex}. The blue wing of \ion{O}{i}$\,\lambda8446$ and the red wing of \ion{Ca}{ii}$\,\lambda$8542 are free from other line contributions. The wings have the same profile as the blue and red wing of \ion{Ca}{ii}$\,\lambda8662$, respectively. We therefore suspect that the \ion{O}{i}$\,\lambda8446$ line and the \ion{Ca}{ii} triplet lines in fact all have the same or at least very similar profiles. To test this assumption, we decompose the \ion{O}{i}\,$\lambda 8446$, \ion{Ca}{ii}\,$\lambda 8498$ and \ion{Ca}{ii}\,$\lambda 8542$ complex using the \ion{Ca}{ii}\,$\lambda 8662$ difference profile as a template for all lines. In fact, we are able to reconstruct the \ion{O}{i}$\,\lambda$8446 and \ion{Ca}{ii}$\,\lambda\lambda$8498, 8542 complex using a \ion{Ca}{ii} triplet ratio of 1:1:1 and an \ion{O}{i}$\,\lambda$8446-to-\ion{Ca}{ii}\,$\lambda 8662$ ratio of 0.85:1. All lines are fixed at their respective central wavelengths ($\pm 50$\,\kms{}). In order to be able to cleanly reconstruct the slope of the blue wing of \ion{O}{i}$\,\lambda$8446, we have to convolve the \ion{Ca}{ii}\,$\lambda 8662$ template with a Gaussian with a width corresponding to $\sigma \approx 70$\,\kms{}.

The reconstruction using three overlapping \ion{Ca}{ii}$\,\lambda$8662 profiles cleanly reproduces the key features of the blend of \ion{O}{i}$\,\lambda8446$ and \ion{Ca}{ii}$\,\lambda\lambda$8498, 8542. In particular, the two peaks in the blended profile can now be clearly attributed to the overlap of \ion{O}{i}$\,\lambda8446$ and \ion{Ca}{ii}$\,\lambda8498$, and of \ion{Ca}{ii}$\,\lambda8498$ and \ion{Ca}{ii}$\,\lambda8542$. Only one larger residuum remains in the overlap between the red and blue wing of \ion{Ca}{ii}$\,\lambda8498$ and \ion{Ca}{ii}$\,\lambda8542$, respectively. The underlying emission of $\sim 0.1\,\times$\,10$^{-15}$ ergs s$^{-1}$cm$^{-2}$\AA$^{-1}$ is on the level of the left continuum after subtraction of the linear pseudo-continuum. We therefore attribute the difference between the reconstructed blended profile and the observed profile to underlying, additional emission not connected to the \ion{O}{i}\,$\lambda 8446$ and \ion{Ca}{ii}\,$\lambda 8498,8542$ complex.

%

%
\subsection{Reconstructing the Balmer line profiles using the \ion{Ca}{ii}\,$\lambda 8662$ difference profile}\label{sec:reconstructed_Balmer_profiles_results}
%
We demonstrate in \ref{sec:OI_CaII_complex_profile_results} that the blended line profile of \ion{O}{i}$\,\lambda8446$ and the \ion{Ca}{ii}$\,\lambda\lambda$8498, 8542 lines can be decomposed into three individual double-peaked profiles closely resembling that of \ion{Ca}{ii}$\,\lambda$8662. While the \ion{O}{i}$\,\lambda8446$ profile and the \ion{Ca}{ii}$\,\lambda\lambda$8498, 8542, 8662 profiles are clearly double-peaked, the Balmer lines lines show no clear indication of double peaks. Instead, the presented rms profiles of \Hb{} and \Ha{} indicate single-peaked emission line profiles, but with a similar redward asymmetry as observed in the \ion{Ca}{ii}$\,\lambda$8662 difference profile. In the \Hb{} and \Ha{} rms profiles, the red wing is $\sim 400$\,\kms{} broader  (with respect to the central wavelength), while the red wing in the \ion{Ca}{ii}$\,\lambda$8662 difference profile is $\sim 300$\,\kms{} broader than the blue wing.

We now show that the Balmer line profiles can be reconstructed from the \ion{Ca}{ii}$\,\lambda8662$ profile by applying a simple broadening function. To this end, we use a three-parameter Lorentzian
\begin{equation}
    I(\lambda)=I_0 \left[\frac{\Gamma^2}{(\lambda-\lambda_0)^2 + \Gamma^2}\right],
\end{equation}
where $\Gamma$ is the half width at half-maximum. This procedure is motivated by the findings of \citet{kollatschny11}, who were able to model turbulent motions in the BLR using Lorentzian line profiles, and of \citet{goad12}, who found that in their bowl-shaped BLR model, Lorentzian line profiles emerge in low-inclination systems for lines formed at larger BLR radii and in the presence of scale-height-dependent turbulence. Therefore, according to the aforementioned models, the broadening by a Lorentzian function in our approach mimics the effects of turbulence in the BLR gas. In addition, other studies also found that emission line profiles in AGN exhibiting line withs of FWHM $\lesssim 4000$\,\kms{} are well approximated by Lorentzian profiles \citep[e.g.,][]{veron-cetty01, sulentic02}.

\begin{figure*}[h!]
    \centering
    \includegraphics[width=1\textwidth, angle=0]{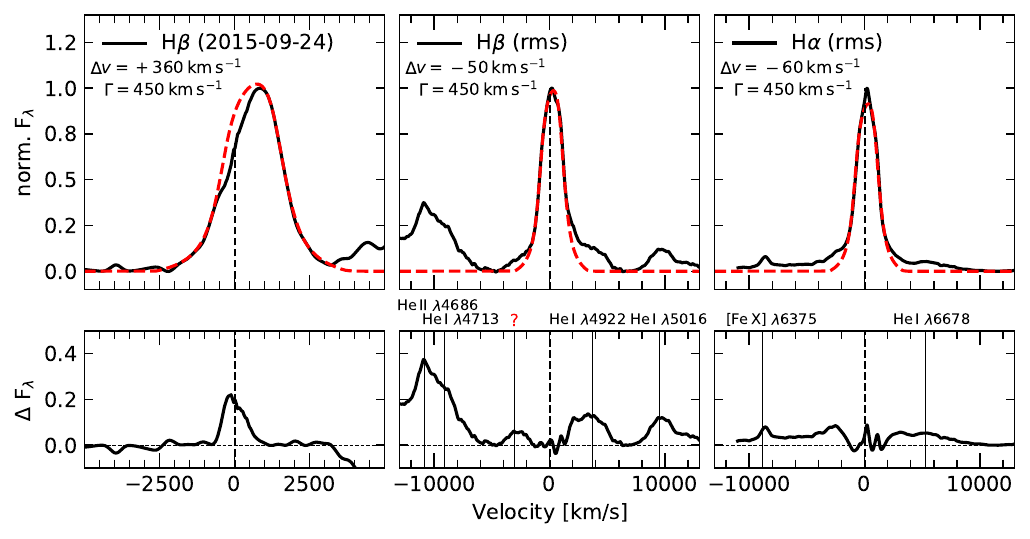}
    \caption{Reconstruction of the \Hb{} and \Ha{} line profiles using the \ion{Ca}{ii}$\,\lambda8662$ line profile, appropriately shifted by a velocity difference $\Delta v$ and broadened with a Lorentzian function of half width $\Gamma$. The resulting smoothed profiles are shown by a dashed red line (\textit{top panels}). The residual fluxes are shown in the \textit{lower panels}. The residual flux of the \Hb{} profile from Spectrum 1 ( 2015 September 24) reveals a small residuum from narrow \Hb{} emission (\textit{lower left panel}). The residual flux of the reconstructed \Hb{} rms profile reveals underlying \ion{He}{i}$\,\lambda4922$ emission and possible emission from an unidentified line species at $4812\,$\AA{} (\textit{lower middle panel}). While the central rms profile of \Ha{} rms is well approximated by a shifted and broadened \ion{Ca}{ii}$\,\lambda8662$ difference profile, we see indications for an additional underlying and extremely broad component in the residual flux (\textit{lower right panel}).}
    \label{fig:Balmer_profile_reconstruction_results}
\end{figure*}

For this procedure, we use the \Hb{} line profile from Spectrum 1 (2015 September 24) as well as the \Hb{} and \Ha{} rms profiles. For each of these profiles, we broaden the \ion{Ca}{ii}$\,\lambda8662$ profile by choosing an appropriate half width $\Gamma$ such that the broadened \ion{Ca}{ii} profile matches the corresponding Balmer line profile. In addition, we introduce a velocity shift $\Delta v$ in order to account for additional blueshifts and redshifts, respectively, of the Balmer lines. The resulting fits and the residual fluxes are shown in Fig.~\ref{fig:Balmer_profile_reconstruction_results}. The \Hb{} line profile from 2015 September 24 is very well approximated by a \ion{Ca}{ii}$\,\lambda8662$ line profile that is shifted by $\Delta v = +360$\kms{} and broadened with a Lorentzian with a  half width of $\Gamma = 450$\,\kms. The only major residuum is a small narrow \Hb{} component. The \Hb{} rms profile is also well approximated by applying Lorentzian broadening of half width $\Gamma = 450$\,\kms{}, but this time with a shift of $\Delta v = -50$\kms{}.In addition to \ion{He}{ii}$\,\lambda4686$, which is already clearly visible in the rms spectrum, we detect emission features at $\sim 4812$\,\AA{}, $\sim 4922$\,\AA{} and $\sim 5016$\,\AA{} in the residual flux in Fig.~\ref{fig:Balmer_profile_reconstruction_results}. Based on the resemblance between the emission features at $\sim 4922,5016$\,\AA{} and the difference line profile of \ion{He}{i}\,$\lambda7065$ (see Fig.~\ref{fig:MUSE_NIR_spectra}), we identify these emission lines as \ion{He}{i}$\,\lambda\lambda4922,5016$. The profiles of \ion{He}{i}$\,\lambda\lambda4922,5016,7065$ as well as of the unidentified emission feature at $\sim 4812$ are analyzed in more detail in \ref{sec:Helium_profiles_results}. The central \Ha{} rms profile is again well approximated by applying Lorentzian broadening of half width $\Gamma = 450$\,\kms{} with a shift of $\Delta v = -60$\kms{}. However, the line wings are less well approximated and residuals between $-2500\,$\kms{} and $+2500\,$\kms{} are clearly visible in the residual flux. In contrast to \Hb{}, the residual flux in \Ha{} indicates an additional and very broad component with a full width at zero intensity (FWZI) of FWZI $\sim 20\,000\,$\kms{}. This is in agreement with the results of \citet{alloin86}, who found that the broad \Ha{} line in \ngc{} line consisted of two components, namely a broad component with an intermediate width of FWHM $=1910\,$\kms{} and a much broader component with a width of FWHM $=6200\,$\kms{}.

%

%
\subsection{Helium line profiles}\label{sec:Helium_profiles_results}
\begin{figure}[h!]
    \centering
    \includegraphics[width=0.47\textwidth, angle=0]{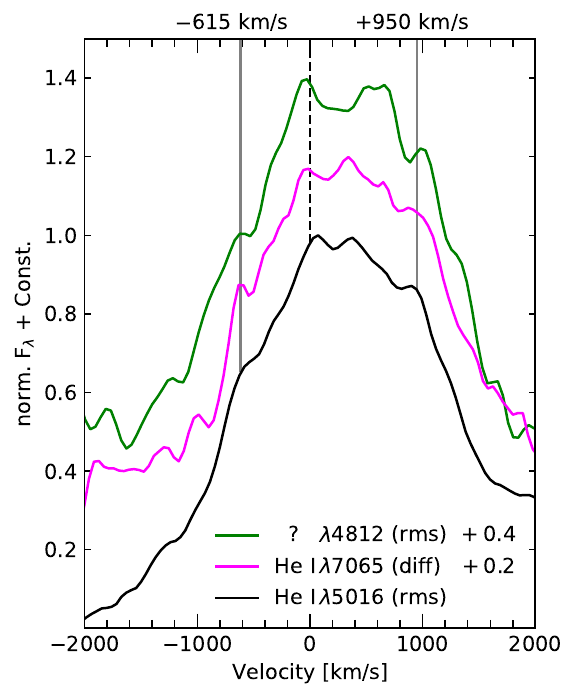}
    \caption{Comparison of the profiles of \ion{He}{i}$\,\lambda5016$, \ion{He}{i}$\,\lambda7065$ and the unidentified line species at approximately $4812\,$\AA{} in velocity space. We subtracted a suitable linear pseudo-continuum from each line and shifted the profiles in flux for clarity. The positions of the left and right peak of \ion{Ca}{ii}$\,\lambda8662$ at $-615$\,\kms{} and $+950$\,\kms{}, respectively, are indicated with gray lines. All three line profiles show indications for a similar double-peaked structure and have an identical width of FWHM $\approx 2170$\,\kms{} when normalized to the profile peak, and of FWHM $\approx 2500$\,\kms{} when the additional central emission is taken into account.}
    \label{fig:Helium_line_profile_results}
\end{figure}

%
We show the rms profiles of \ion{He}{i}$\,\lambda5016$ as well as of the unidentified emission at $4812\,$\AA{}, and the difference line profile of \ion{He}{i}$\,\lambda7065$ in Fig.~\ref{fig:Helium_line_profile_results}. In each case, we subtracted a suitable linear pseudo-continuum. All three emission lines have an identical width of FWHM $\approx 2170$\,\kms{} and show indications of a double-peak structure with the red peak again being stronger than the blue peak. The peaks are positioned at $-615$\,\kms{} and $+950$\,\kms{}, and are therefore identical to the peak positions in \ion{Ca}{ii}$\,\lambda8662$ (see \ref{sec:CaII_profile_results}). In addition, we again observe a redward asymmetry with the red wing being broader by $~\sim 400$\,\kms{} with respect to the central velocity.

In contrast to \ion{Ca}{ii}$\,\lambda8662$ (see Fig.~\ref{fig:CaII_8662_veloplot}), the three profiles (see Fig.~\ref{fig:Helium_line_profile_results}) do not exhibit a central and skewed dip, but instead show additional emission in the region $-450\,$\kms{} $< v < +550\,$\kms{}. Supposing that this additional component is superimposed on a double-peaked profile similar to that of \ion{Ca}{ii}$\,\lambda8662$, the FWHM of the genuine double-peaked \ion{He}{i}$\,\lambda\lambda5016,7065$ without the additional component can be estimated to be closer to FWHM $\sim 2500$\,\kms{}.
%

%
\subsection{The \ion{H}{II} region close to the nucleus}\label{sec:HII_region_results}
\begin{figure*}[h!]
    \centering
    \includegraphics[width=1\textwidth,angle=0]{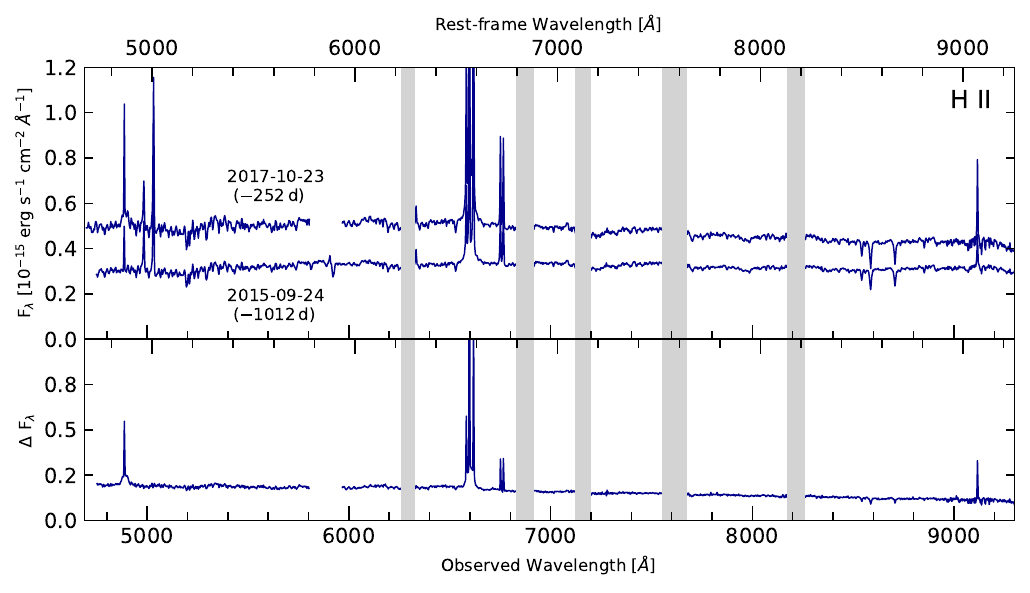}
    \caption{Spectrum of the \ion{H}{ii} region at a distance of $\Delta d \sim$1\arc0 from the nucleus taken on 2015 September 24 and 2017 October 23, respectively, intercalibrated to the same [\ion{O}{iii}]\,$\lambda5007$ flux of $(4.2 \pm 0.3)\, \times$\,10$^{-15}$ ergs s$^{-1}$ cm$^{-2}$ (\textit{upper panel}) and the resulting difference spectrum (\textit{lower panel}). Spectral regions strongly influenced by telluric absorption are shown in gray.  The apparent continuum flux increase on 2017 October 23 is caused by scattered radiation from the brightening nucleus.}
    \label{fig:HII_region_spectra}
\end{figure*}

%
An \ion{H}{ii} region at a separation of only $\sim 1$\arc0 (corresponding to $\sim 100$\,pc) from the nucleus of \ngc{} was detected by \citet{smajic15} and \citet{silva18}. Due to the close proximity to the nucleus, we investigate the possible extent of contamination from this region in the nuclear spectra. For this purpose, we extract the spectrum of the \ion{H}{II} region from the MUSE data cubes from 2015 September 24 and 2017 October 23, respectively, using a circular aperture of 0\arc5. Due to the favorable seeing conditions on 2017 October 23 and the spectrum being taken in AO wide field mode, we scale the spectrum from 2015 September 24 to the integrated [\ion{O}{iii}]\,$\lambda5007$ flux of $(4.2 \pm 0.3)\, \times$\, 10$^{-15}$ ergs s$^{-1}$ cm$^{-2}$ from 2017 October 23. The spectra are shown in Fig.~\ref{fig:HII_region_spectra} together with the resulting difference spectrum. Both spectra are dominated by narrow line emission of \Hb\,$\lambda$4861, [\ion{O}{iii}]\,$\lambda\lambda$4959, 5007, [\ion{N}{ii}]\,$\lambda\lambda$6548, 6583, \Ha\,$\lambda$6563, [\ion{S}{ii}]\,$\lambda\lambda$6716, 6731, [\ion{S}{iii}]\,$\lambda$9069, absorption of the \ion{Ca}{ii} triplet, and a strong underlying stellar contribution. For the low-state spectrum from 2015 September 24, the V-band flux amounts to $\sim 0.3\, \times$\,10$^{-15}$ ergs s$^{-1}$ cm$^{-2}$ \AA{}$^{-1}$.
The spectra reveal an apparent increase in flux from 2015 September 24 to 2017 October 23, both in the most prominent narrow lines except for [\ion{O}{iii}]\,$\lambda\lambda$4959, 5007 as well as in the underlying continuum emission. We attribute this to narrow-line flux losses in the extended narrow-line region (NLR) due to unfavorable seeing conditions on 2015 September 24 on the one hand, and scattered radiation from the brightening nucleus on 2017 October 23 on the other hand. The effect of the stray emission from the brightening nucleus is best seen in the \Hb{} line, where a small broad component appears to emerge on 2017 October 23.

%
\section{Discussion}\label{sec:discussion}
%

%
\subsection{Influence of the \ion{H}{ii} region on the AGN spectra}\label{sec:HII_region_discussion}
%
Stray emission from the \ion{H}{ii} region at a distance of $\Delta d \sim$1\arc0 from the nucleus can in principle effect the observed AGN spectra by contributing additional continuum and line flux, thereby also effecting the intercalibration of the AGN spectra on the basis of the [\ion{O}{iii}]\,$\lambda5007$ line. In order to assess the potential effect of stray emission, we inspect the MUSE \ion{H}{ii} region spectra from 2015 September 24 and 2017 October 24 in \ref{sec:HII_region_results} and determine an integrated [\ion{O}{iii}]\,$\lambda5007$ flux of $(4.2 \pm 0.3)\, \times$\, 10$^{-15}$ ergs s$^{-1}$ cm$^{-2}$. This is 4\% of the integrated [\ion{O}{iii}]\,$\lambda5007$ of the nuclear region. However, since the extraction apertures for the nuclear region and the \ion{H}{ii} region in the MUSE spectra were chosen such that the overlap between the apertures is minimal (see \ref{sec:MUSE_observations}), the real contribution from additional [\ion{O}{iii}]\,$\lambda5007$ from the \ion{H}{ii} region can be securely estimated to be $<1$\%, even when taking into account seeing effects. For the SALT spectra, the square aperture of 2\arc0 $\times$ 2\arc0 for the nuclear region slightly increases the contribution from the \ion{H}{ii} region, and modest inaccuracies in the exact slit pointing might add to this effect. Nevertheless, the additional contribution in [\ion{O}{iii}]\,$\lambda5007$ flux can conservatively be estimated to be $<2$\%.

From the MUSE spectra of the nuclear region and the \ion{H}{ii} region taken on 2015 September 24, we find that the V-band flux in the \ion{H}{ii} region is 7\% of that in the low-state nuclear spectrum. However, this measurement still includes the strong stellar host-contribution, which we estimate to account for $\sim 70$\% of the V-band flux (see \ref{sec:host_galaxy_contribution_results}). This reduces the potential continuum contribution of the \ion{H}{ii} region to the host-free nuclear spectra to $\sim 4$\%. Taking into account the minimal overlap between apertures, this further reduces the contribution to $<1$\% for the MUSE and $<2$\% for the SALT spectra. We conclude that the \ion{H}{ii} region contributes only insignificantly to the nuclear spectra presented in this paper.

%
\subsection{The double-peaked \ion{Ca}{ii} triplet and \ion{O}{i} line profiles}\label{sec:double_peaked_profiles_discussion}
%

%
\subsubsection{Robustness of the double-peaked difference profiles}\label{sec:robustness_line_profile_discussion}
%
We show in \ref{sec:CaII_profile_results} that the difference line profile of \ion{Ca}{ii}\,$\lambda8662$ is ``lopsided'' and double-peaked, exhibiting a skewed dip in the central profile. It closely resembles line profiles observed in a number of AGN and TDEs  (see \ref{sec:CaII_profile_results}), which are interpreted to arise in an elliptical accretion disk. To our knowledge, this is the first time ``genuine'' double-peaked \ion{Ca}{ii} triplet emission -- as well as double-peaked $L\beta$-pumped \ion{O}{i}$\,\lambda8446$ emission -- in AGN has been presented in the literature. Only recently, \citet{diasdossantos23} found, for the first time, a double-peaked \ion{O}{i}$\,\lambda11279$ profile in III Zw 002. Out of the 14 \ion{Ca}{ii} emitters shown by \citet{persson88b}, none shows unambiguous indications for a double-peaked profile caused by gas kinematics. Instead, \citet{persson88b} attributes the central dips present in some \ion{Ca}{ii} emitters, such as Mrk\,42, to underlying stellar absorption from the host galaxy. However, genuine double-peaked \ion{Ca}{ii} profiles were detected in spectra of cataclysmic variables and associated with emission from an accretion disk \citep[e.g.,][and references therein]{persson88a}.

We omit the effect of \ion{Ca}{ii} absorption by extracting the \ion{Ca}{ii} difference line profile between the observations taken on 2015 September 24 and 2017 October 23, respectively. The robustness of the double-peaked profile therefore depends directly on the quality of the intercalibration of these two spectra. The quality of the intercalibration can be directly assessed from Fig.~\ref{fig:MUSE_NIR_spectra}. The two spectra we intercalibrated such that the stellar signature of the host galaxy and the narrow lines in the difference profile vanish. The difference spectrum reveals only broad line emission, indicating a successful intercalibration of the two spectra. 

\begin{figure}[h!]
    \centering
    \includegraphics[width=0.47\textwidth,angle=0]{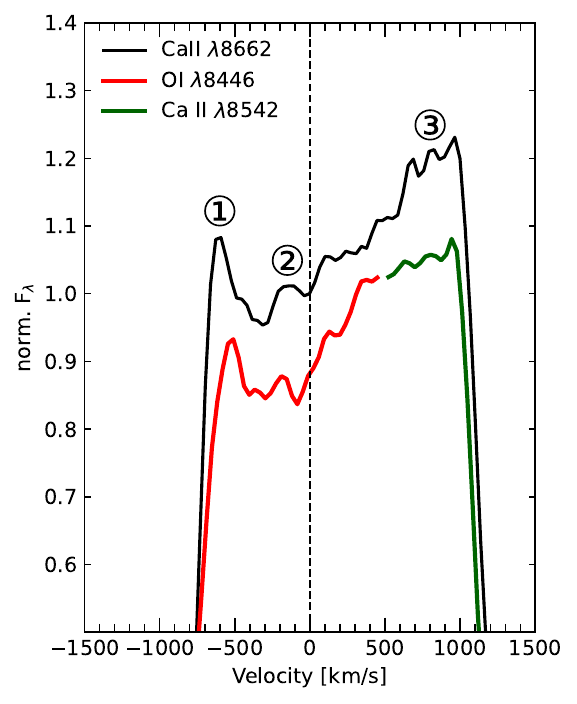}
    \caption{Comparison of the \ion{Ca}{ii}$\,\lambda8662$ difference line profile (\textit{black}) with the blue wing of \ion{O}{i}$\,\lambda8446$ (\textit{red}) and the right wing of \ion{Ca}{ii}$\,\lambda8542$ (\textit{green}) in velocity space. The profiles of \ion{O}{i}$\,\lambda8446$ (\textit{red}) and \ion{Ca}{ii}$\,\lambda8542$ are normalized to their left and right peak, respectively, and shifted downwards for clarity.  All profiles exhibit identical line features, namely a pronounced left peak \raisebox{.5pt}{\textcircled{\raisebox{-.9pt} {1}}}, the same lopsidedness and characteristic small-scale features in the central profile \raisebox{.5pt}{\textcircled{\raisebox{-.9pt} {2}}}, and a triple-peaked (trident) red peak \raisebox{.5pt}{\textcircled{\raisebox{-.9pt} {3}}}.}
    \label{fig:CaII_OI_comparison}
\end{figure}

The strongest argument for a successful correction for \ion{Ca}{ii} absorption is provided by comparing the line profile of \ion{Ca}{ii}\,$\lambda8662$ with that of \ion{O}{i}$\,\lambda8446$ and \ion{Ca}{ii}\,$\lambda8542$. The blue wing of
\ion{O}{i}$\,\lambda8446$ is neither influenced by strong absorption nor by blending with \ion{Ca}{ii}\,$\lambda8498$. Likewise, the red wing of \ion{Ca}{ii}\,$\lambda8542$ is free of absorption as well as line blending effects. We show a comparison of the \ion{Ca}{ii}$\,\lambda8662$ difference line profile with the blue wing of \ion{O}{i}$\,\lambda8446$ and the right wing of \ion{Ca}{ii}$\,\lambda8542$ in Fig.~\ref{fig:CaII_OI_comparison}. The profiles of \ion{O}{i}$\,\lambda8446$ and \ion{Ca}{ii}$\,\lambda8542$ are normalized to their left and right peak, respectively, and shifted downwards for clarity. All profiles exhibit identical line features, namely a pronounced left peak \raisebox{.5pt}{\textcircled{\raisebox{-.9pt} {1}}}, the same lopsidedness and characteristic small-scale features in the central profile \raisebox{.5pt}{\textcircled{\raisebox{-.9pt} {2}}}, and a triple-peaked red peak \raisebox{.5pt}{\textcircled{\raisebox{-.9pt} {3}}} that resembles a trident. Although the exact position, width and scaling of the blue peak and the small-scale features in the \ion{O}{i}$\,\lambda8446$ profile differ slightly from that in the \ion{Ca}{ii}$\,\lambda8662$ profile, the qualitative profile is identical.  We therefore conclude that the \ion{Ca}{ii}$\,\lambda8662$ difference line profile presented in Fig.~\ref{fig:CaII_8662_veloplot} is in fact the clean profile and that residual \ion{Ca}{ii} absorption is not significantly affecting the line profile. More precisely, the skewed, lopsided profile is exclusively due to the structure and kinematics of the BLR.

%

%
\subsubsection{The elliptical accretion disk model for \ion{Ca}{ii}$\,\lambda8662$}\label{sec:disk_origin_discussion}
%
We show in \ref{sec:fitting_results} that the \ion{Ca}{ii}\,$\lambda8662$ difference profile (and in turn the \ion{O}{i} profile and the other \ion{Ca}{ii} profiles), with exception of the inner part of the profile, is in general well approximated by line emission from a relativistic elliptical accretion disk. In particular, the model reproduces the observed shape of the line wings, and only the red peak of the model fit is marginally shifted inwards by about 50\,\kms{} with respect to the observed profile. This indicates that the structure (and/or kinematics) of the disk generating the line profile might be more complex than a homogeneous elliptical disk, thereby causing the trident structure in the red peak (see \ref{sec:disk_inhomogeneities_discussion} for further details).

The model reproduces a low disk inclination of $i = (8.10 \pm 3.00)^\circ$ in agreement with findings of \citet{parker19}, who found $i < 11^{\circ}$, and with the almost face-on view of the host galaxy, though we note that \ngc{}'s host-galaxy geometry is more complex on larger scales \citep{elagali19}. The major axis viewing angle, which is the angle between the major axis in apocenter direction and the projected line-of-sight in the disk plane, is $\phi_0 = (193.29 \pm 26.00)^{\circ}$. The disk is confined within the inner and outer pericenter distances $\xi_1$ and $\xi_2$ of $(2231 \pm 1000)$\,r$_g$ and $(4050 \pm 1500)$\,r$_g$, respectively, and exhibits very low internal broadening with $\sigma = (87 \pm 10)$\,\kms{}. The eccentricity of $e = 0.57 \pm 0.35$ is moderate, and the emissivity power-law index $q$ amounts to $4.34 \pm 0.80$.  This is larger than the value of $q \approx 3$ for the emissivity profile of an outer accretion disk ($\gtrsim 100$\,r$_g$) irradiated by an isotropic point source (or simple extended source) of X-ray emission \citep[see][and references therein]{wilkins12} usually adopted in other studies \citep[e.g.,][]{eracleous94, eracleous03}.

Our best-fit model is able to reproduce all key features of the \ion{Ca}{ii}\,$\lambda8662$ profile, namely the redward assymetry and the narrow blue and red peak, respectively. It cannot, however, account for all of the emission in the central part of the profile between $-450\,$\kms{} $< v < +850\,$\kms{}. This discrepancy between disk-modeled line profiles and observed line profiles has been noticed in other studies \citep[e.g.,][]{hung20, wevers22, diasdossantos23}, and has been attributed to additional emission from gas above the disk plane. In contrast to previous studies, we do not attempt to recover the additional emission component by a Gaussian component as this component in our model is clearly not a Gaussian. Instead, the central emission component has a distorted, rather flat-topped profile with a pronounced redward asymmetry.

In agreement with the aforementioned studies, we attribute this additional emission component to emission from BLR clouds situated above the accretion disk plane, apparently not retaining the angular momentum of the disk. The observed redward asymmetry might be explained by the asymmetric distribution of clouds with respect to the ionizing continuum, which we assume to be in close vicinity to the SMBH. Regardless of the \textit{exact} distribution of BLR gas above the disk, we conclude that the BLR in \ngc{} is a two-component BLR, consisting of a BLR disk component and an additional component of BLR gas above the disk plane.

For a sample of six AGN, \citet{eracleous95} found pericenter distances $\xi_1$ and $\xi_2$ for \Ha{} ranging between $350 - 1900$\,r$_g$ and $3000 - 9000$\,r$_g$, respectively. For a sample of 116 double-peaked Balmer line AGN, \citet{strateva03} found the accretion disks to be consistent with inner radii of ($200-800)$\,r$_g$, and outer radii $\geq 2000$\,r$_g$. This places our inner pericenter distance $\xi_1$ at the higher end of the distribution, and the outer pericenter distance $\xi_2$ at the lower end. Nevertheless, the scatter in the modeled disk extents for different objects is rather large. For example, \citet{ricci19} found both pericenter distances $\xi_1$ and $\xi_2$ for \Ha{} in NGC\,4958 to be $< 1000$\,r$_g$. \citet{storchi-bergmann97} found pericenter distances $\xi_1$ and $\xi_2$ of $1300\,$r$_g$ and $1600\,$r$_g$ in NGC 1097 using their refined elliptical disk model, indicating a rather ring-like structure. 

In order to test the robustness of our solution, we now calculate the expected lag of \Hb{} using the R$_{H\beta} - \lambda$L$_{\rm 5100}$ relationship
\begin{equation}
    {\rm log}\left[\frac{R}{\rm 1 lt-day}\right] = K + \beta\, {\rm log}\left[\frac{\lambda L_{5100}}{10^{44} {\rm erg s}^{-1}} \right],
\end{equation}
where K is the origin and $\beta$ is the slope of the relation.  We expect \Hb{} to be emitted at approximately the same radial distance as \ion{Ca}{ii}$\,\lambda8662$, since we can model the \Hb{} profile as a \ion{Ca}{ii}$\,\lambda8662$ profile broadened by a Lorentzian function. This is interpreted as the effect of scale-height-dependent turbulence (see \ref{sec:turbulence_and_stratification_discussion}), which in turn means that the underlying rotational kinematics of \Hb{} and \ion{Ca}{ii} are identical. Recent works suggest that the FWHM of \Hb{} and \ion{Ca}{ii} in high-luminosity sources are tightly correlated, while the FWHM of \Hb{} in local low-luminosity sources like \ngc{} is slightly larger than that of \ion{Ca}{ii} \citep{martinez-aldama21}. We use a slope of $\beta = 0.53$ as found by \citet{bentz13} and \citet{kollatschny18}, $K = 1.53$, and adopt $\lambda$L$_{\rm 5100} = 3.91 \times 10^{41}$\, ergs s$^{-1}$ obtained in \ref{sec:black_hole_mass_results}. This gives a BLR radius of only $\sim 1.8$\,lt-days. We now compare this value with the delay expected from the best-fit elliptical accretion disk model calculated in \ref{sec:fitting_results}. For this purpose, we first calculate the distance $d =\xi_2 / (1-e) - \xi_2$ of the center of the disk to the SMBH using the outer perigee $\xi_2$ of the disk emitting region and the eccentricity $e$. This gives a distance of $d \approx 5500$\,r$_g$ from the SMBH to the disk center, and with M$_{\rm BH} = (5.3 \pm 2.7) \times 10^6$\,M$_{\odot}$ calculated in \ref{sec:black_hole_mass_results}, which results in a gravitational radius r$_g = GM/c^2$ of $0.0003$\,lt-days, we obtain $\sim 1.7$\,lt-days for the expected delay between continuum and line emission in \ngc{}. This is a rough estimate, since the line emission is not weighted by emissivity. Nonetheless, this simple order-of-magnitude assessment confirms that the best-fit disk parameters are in good agreement with established reverberation mapping scaling relations.

%
\subsubsection{The Bowen fluorescence line \ion{O}{i}$\,\lambda8446$ as part of the elliptical disk model}\label{sec:Bowen_fluorescence_discussion}
\begin{figure}[h!]
    \centering
    \includegraphics[width=0.47\textwidth,angle=0]{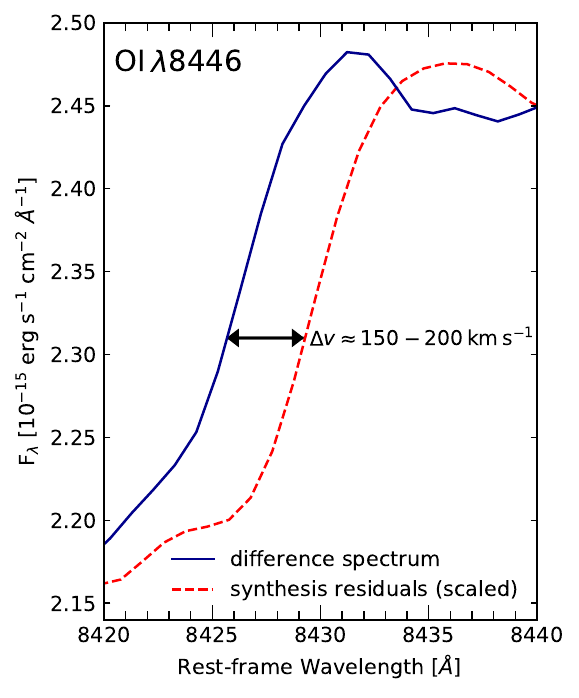}
    \caption{Comparison of the blue wing of \ion{O}{i}$\,\lambda8446$ between the difference spectrum from 2015 September 24 to 2017 October 23 (\textit{blue}) and the host-subtracted residuum spectrum from 2015 September 24 (slightly smoothed and scaled for clarity; \textit{red}) resulting from the stellar synthesis in \ref{sec:host_galaxy_contribution_results}. The shift of the blue wing of \ion{O}{i}$\,\lambda8446$ from 2015 September 24 to 2017 October 23 indicates a blueward drift of the line emission. The difference velocity $\Delta v_{\rm shift}$ amounts to $\Delta v_{\rm shift} \approx 150-200$\,\kms{}.}
    \label{fig:comparison_OI_OI}
\end{figure}

%
We find indications for \ion{O}{i}$\,\lambda8446$ as well as \ion{Ca}{ii} triplet emission already being present during the low-state, host-galaxy subtracted spectrum from 2015 September 24 (Spectrum 1; see Fig.~\ref{fig:spectral_synthesis}). Although the S/N of \ion{O}{i} and \ion{Ca}{ii} in the host-galaxy subtracted spectrum is low and the \ion{Ca}{ii} triplet is still affected by residual absorption, we identify the blue wing of \ion{O}{i}$\,\lambda8446$ together with its blue peak. We show a comparison between the blue \ion{O}{i}$\,\lambda8446$ wing from 2015 September 24 and the blue \ion{O}{i}$\,\lambda8446$ wing from the difference spectrum in Fig.~\ref{fig:comparison_OI_OI}. The blue wing of \ion{O}{i}$\,\lambda8446$ from the low-state spectrum is scaled and smoothed for clarity. We find a blueward drift of about $- (150 - 200)$\,\kms{} with respect to the low-state spectrum. For comparison, \Hb{} shifted by $-240$\,\kms{} between Spectrum 1 and Spectrum 2 (from 2015 September 24 to 2017 October 23). This indicates that \ion{O}{i}$\,\lambda8446$ shifts in tandem with the Balmer lines (see \ref{sec:optical_profiles_results}), but not necessarily by the exact same amount.\footnote{We also find apparent blueward drifts of the same order in the red wings of \ion{Ca}{ii}$\,\lambda8662$ and \ion{Ca}{ii}$\,\lambda8662$. However, the S/N in the host-galaxy-subtracted spectrum is too low and the profiles are affected by residual absorption, meaning that no robust quantitative estimate of the shift can be made.} 

In the following, we discuss the production channels of \ion{O}{i}$\,\lambda8446$ and its region of origin. \ion{O}{i}$\,\lambda8446$ emission can be produced by either recombination, collisional excitation by electrons, continuum fluorescence, Ly$\beta$ fluorescence (a process first described by \citet{bowen47}, hence also termed Bowen fluorescence) or a combination of these mechanisms \citep[for a review see][]{grandi80}. The exact mechanism(s) responsible for the emission of \ion{O}{i}$\,\lambda8446$ in a particular source can be determined by performing line diagnostics with respect to other \ion{O}{i} lines in this source. \citet{rodriguez-ardila02b} performed line diagnostics on the \ion{O}{i} lines $\lambda\lambda1304,7774,8446,11287$ in a sample of six AGN. They found \ion{O}{i}$\,\lambda8446$ to be formed through collisional excitation and Ly$\beta$ fluorescence, and excluded recombination and continuum fluorescence as a source of the observed line strength of \ion{O}{i}$\,\lambda8446$. \citet{matsuoka07} later corroborated these results by means of photoionization models. 

\ion{O}{i}$\,\lambda8446$ emission in AGN has been known for several decades, and has early been associated with the broad line region due to the absence of a narrow component \citep{grandi80}. We can confirm the absence of a narrow \ion{O}{i}$\,\lambda8446$ component in \ngc{}, and, moreover, report that \ion{O}{i}$\,\lambda8446$ and the lines of the NIR \ion{Ca}{ii} triplet exhibit virtually identical line profiles (see Fig.~\ref{fig:CaII_OI_comparison}). \citet{persson88b} found a tight correlation between the FWHM of \ion{O}{i}$\,\lambda8446$ and the \ion{Ca}{ii} triplet, indicating that the \ion{O}{i} and \ion{Ca}{ii} lines stem from approximately the same region. Likewise, \citet{rodriguez-ardila02a} concluded by virtue of identical line profiles, that \ion{O}{i} and \ion{Ca}{ii} share the same kinematics. We can confirm this result based on the identical line profiles, that place the emitting gas in a region dominated by the kinematics of an elliptical disk (see \ref{sec:disk_origin_discussion}). We stress, however, that the profile of \ion{O}{i}$\,\lambda8446$ exhibits a more redshifted blue peak and a shallower blue wing than \ion{Ca}{ii}\,$\lambda8662$. This required convolving the \ion{Ca}{ii}\,$\lambda8662$ profile with a Gaussian of small width in order to reproduce the slope of the left wing of \ion{O}{i}$\,\lambda8446$ when reconstructing the blended \ion{O}{i} and \ion{Ca}{ii} triplet complex. Such a slight change in the profile's shape can be achieved in the best-fit \ion{Ca}{ii}\,$\lambda8662$ model by increasing the inner pericenter distance $\xi_1$ by about 10\%. We therefore speculate that emission of \ion{O}{i}$\,\lambda8446$ takes place in a similar, overlapping region with respect to emission of \ion{Ca}{ii}, and that this region has a marginally larger inner radius.

Bowen fluorescence lines have recently gained attention by the detection and definition of a new class of transient events in AGN by \citet{trakhtenbrot19a}. They described a new class of long-lived (a few years) transient events during which broad \ion{N}{iii}$\,\lambda4640$ \citep[as well as \ion{O}{iii}$\,\lambda3133$ and \ion{N}{iii}$\lambda\lambda4097,4103$; see][and references therein]{makrygianni23} is generated by the absorption of \ion{He}{ii} Ly$\alpha$ by \ion{O}{iii}. \ion{O}{iii} then  de-excites through a series of optical transitions and a far-UV transition at $374.436$\,\AA{}, which in turn excites \ion{N}{iii}, generating optical emission of \ion{N}{iii}$\,\lambda\lambda4097,4104,4379,4634,4641$ \citep{bowen28, bowen35}. We note that L$\beta$ pumping, and therefore the generation of \ion{O}{i}$\,\lambda8446$, follows a different production channel, namely the absorption of \ion{H}{i} L$\beta$ ($1025.72,\AA{}$) by \ion{O}{i}, directly leading to the emission of \ion{O}{i}$\,\lambda\lambda1304,8446,11287$ through optical de-excitation.

The presence of enhanced ionizing UV emission during Bowen fluorescence flares has been attributed to enhanced accretion onto the SMBH \citep{trakhtenbrot19a, makrygianni23} in a pre-existing AGN. Based on the detection of weak \ion{O}{I}$\,\lambda8446$ emission in \ngc{} present already before the transient event, we likewise attribute the enhanced UV emission \citep{oknyansky20} and the in turn enhanced \ion{O}{I}$\,\lambda8446$ emission during the event to an increase in accretion rate.

%
\subsubsection{Density and column density in the \ion{O}{i} and \ion{Ca}{ii} disk}\label{sec:column_density_BLR_disk_discussion}
%
Based on the almost identical line profiles of \ion{O}{i}$\,\lambda8446$ and \ion{Ca}{ii}, we show in \ref{sec:Bowen_fluorescence_discussion} that both lines originate from approximately the same region, and that this region is dominated by the kinematics of an elliptical accretion disk. We now show that this is consistent with photoionization calculations, which indicate that both \ion{O}{i}$\,\lambda8446$ and \ion{Ca}{ii} are emitted from a region of similar (column) density. In \ref{sec:OI_CaII_complex_profile_results}, we find a \ion{Ca}{ii} triplet ratio of 1:1:1 and an \ion{O}{i}$\,\lambda8446$-to-\ion{Ca}{ii}$\,\lambda8662$ ratio of 0.85:1 . The \ion{Ca}{ii} T : \ion{O}{i}$\,\lambda8446$ ratio is therefore 3.5. \citet{persson88b} pointed out that if \ion{Ca}{ii} triplet emission is detected in AGN, the \ion{Ca}{ii} triplet ratio is usually on the order of 1:1:1. This indicates that the \ion{Ca}{ii} emission arises from a region that is optically thick in \ion{Ca}{ii}. Model calculations for \ion{Ca}{ii} \citep{joly89, ferland89} placed the emitting region in a cool and dense gas with a temperature $T \leq 8000$\,K, a density of $n_H \simeq 10^{12}$\,cm$^{-3}$, and a column density of $N_H \geq 10^{23}$\,cm$^{-2}$. More recent studies corroborated the results for the gas and column density \citep{panda20,panda21a,panda21b}. These physical conditions are the conditions of an outer, cold accretion disk, and \citet{ferland89} and later \citet{dultzin-hacyan99} associated \ion{Ca}{ii} emission with a wind (or corona) just above the accretion disk. Based on photoionization model calculations on surveys of intermediate-redshift quasars, \citet{matsuoka07} and \citet{martinez-aldama15} suggested that \ion{O}{i}$\,\lambda8446$ and the \ion{Ca}{ii} triplet are emitted from regions with similar physical conditions. In particular,  \citet{matsuoka07} restricted the density of the \ion{O}{i} and \ion{Ca}{ii} emitting gas to $n_H \simeq 10^{11.5}$\,cm$^{-3}$, and based on our measured \ion{Ca}{ii} T : \ion{O}{i}$\,\lambda8446$ ratio of 3.5 in \ngc{}, their model gives a column density of $N_H \simeq 1.2 \times 10^{23}$\,cm$^{-2}$ \citep[see Fig.~8 in][]{matsuoka07} for the \ion{O}{i}$\,\lambda8446$ emitting region in \ngc{}. In summary, we suggest that \ion{O}{i} and \ion{Ca}{ii} are emitted in  a similar, overlapping region, as argued before on basis of the identical line profiles.

%
\subsection{Helium profiles and the unidentified emission at $4812\,$\AA{}}\label{sec:Helium_discussion}
%
In \ref{sec:reconstructed_Balmer_profiles_results} and \ref{sec:Helium_profiles_results} we identify emission of \ion{He}{i}$\,\lambda\lambda4922,5016,7065$ based on their identical line profiles. The profiles of \ion{He}{i}$\,\lambda\lambda5016,7065$ show, in comparison to the profiles of \ion{O}{i}$\,\lambda8446$, \ion{Ca}{ii}$\,\lambda\lambda8498, 8542, 8662$, and the ``smoothed'' profile of \Hb{}, a much stronger central emission component, whereas the double-peak structure is almost completely suppressed. The \ion{He}{i} profiles bear a remarkable resemblance to the late-time \Ha{} profiles observed in the TDE AT 2018hyz by \citet{hung20}, when the central additional emission starts to dominate the line profile. \citet{hung20} identified the additional emission to be a nondisk Gaussian component related to a BLR that formed by radiatively driven winds. Likewise, the BLR in \ngc{} can be modeled as a two-component BLR, namely BLR disk and BLR gas above the disk plane, where the BLR component from above the disk contributes more to the \ion{He}{i} lines than to the other line species. At the same time, the FWHM of the \ion{He}{i} disk-component in \ngc{} is $\sim 2500$\,\kms{} and therefore broader  by $\sim 600$\,\kms{} and $\sim 300$\,\kms{} in comparison to \ion{Ca}{ii}$\,\lambda8662$ and \Hb{}, respectively. 

The detection of the yet unidentified emission at $\sim 4812\,$\AA{} is based on the striking resemblance between this emission line and the \ion{He}{i}$\,\lambda\lambda5016,7065$ profiles, all of them exhibiting approximately the same width, weak double-peak features and additional strong emission in the central line profile with almost identical small-scale features. The pronounced similarity between the profiles might suggest that this is genuine emission and not just a residual from the reconstruction of \Hb{} from the double-peaked \ion{Ca}{ii} profile. If indeed line emission, the exact identification of the line species is difficult, since the central wavelength of $\sim 4812\,$\AA{} is estimated by comparing the line profile with that of \ion{He}{i}$\,\lambda\lambda5016,7065$. As shown previously, \Hb{} and \ion{O}{i}$\,\lambda8446$ show significant, but not necessarily identical blueward drifts during the rising phase of the transient event, and we suspect the same to be true for all broad lines. Consequently, we cannot exactly determine the central wavelength of the unidentified emission. Despite the resemblance to the \ion{He}{i} profiles, we find the emission at $\sim 4812$\,\AA{} unlikely to be associated with \ion{He}{i} as no emission from \ion{He}{i} is expected in a spectral range $\pm 100$\,\AA{} around this wavelength (see the NIST Atomic Spectra Database\footnote{\url{https://physics.nist.gov/PhysRefData/ASD/lines_form.html}}).

\citet{veron02} associated the ``red shelf'' observed in some Seyfert 1s \citep{meyers85} with broad emission from \ion{He}{i}$\,\lambda4922$ and \ion{He}{i}$\,\lambda5016$. We confirm the presence of \ion{He}{i}$\,\lambda\lambda4922,5016$ in \ngc{} and that they constitute the red shelf of \Hb{} clearly discernible in the rms spectrum close to \Hb{} (see Fig.~\ref{fig:Balmer_profile_reconstruction_results}). Although the blue wing of \ion{He}{i}$\,\lambda4922$ is slightly disturbed by either interference from small \Hb{} residuals or unidentified line emission, the red wing shows the same shape as the other \ion{He}{i} lines. We also find indications for the red wing of \ion{He}{i}$\,\lambda4713$, however, this line is heavily blended with \ion{He}{ii}$\,\lambda4686$. The spectral region between $\sim 4500$\,\AA{} and $\sim 5600$\,\AA{} in AGN can be densely populated by line emission of \ion{Fe}{ii}, and previous studies found tight correlations between the properties of \ion{Ca}{ii} and \ion{Fe}{ii} emitting regions \citep{panda20, panda21a, panda21b}. In particular, \ion{Ca}{ii} and \ion{Fe}{ii} are expected to be emitted from gas with the same temperature, density, column density, and their line profile widths indicate an origin from approximately the same spatial region \citep{marinello16, marinello20}. Hence, since \ngc{} shows strong \ion{Ca}{ii} emission, significant optical \ion{Fe}{ii} emission is expected. Indeed, we observe strong line emission between $\sim 4500$\,\AA{} and $\sim  5600$\,\AA{} during the transient event, requiring a careful decomposition and fitting in order to determine the different line contributions. This is, however, beyond the scope of this paper.

%
\subsection{Broad-line region structure and kinematics}\label{sec:BLR_structure_discussion}
%

%
\subsubsection{The role of turbulence and stratification in the BLR}\label{sec:turbulence_and_stratification_discussion}
%
Spectrum 1 (2015 September 24) and Spectrum 2 (2017 October 23) exhibit very similar \Hb{} line profiles, and we show in  \ref{sec:reconstructed_Balmer_profiles_results} that the low-state \Hb{} profile is in perfect agreement with a double-peaked \ion{Ca}{ii}$\,\lambda8662$ profile smoothed by a Lorentzian of  half width $\Gamma =450$\,\kms{}. The resulting profile itself is not a Lorentzian profile, but is asymmetrical and slightly skewed due to the underlying double-peaked profile. Likewise, we can model the rms profiles of \Hb{} and \Ha{} by exactly the same procedure.\footnote{We note that the rms spectrum is dominated by the two strongest spectra closest to the transient peak.} This indicates that underlying kinematics do not change drastically during the transient event. We draw two major conclusions from these results: First, since the low-state \Hb{} profile before the transient event is evidently shaped by the same elliptical-disk kinematics as \ion{Ca}{ii}$\,\lambda8662$, the disk-like BLR was already present before the transient event began in 2017. Second, the \Hb{} profile is influenced by some physical process effectively smearing out the profile with respect to the double-peaked \ion{Ca}{ii}$\,\lambda$ profile. In the following, we show that this smearing of the profile can be attributed to turbulence.

\citet{kollatschny13a} and \citet{kollatschny13b} were able to parameterize broad emission lines in AGN by the ratio of FWHM to line dispersion $\sigma_{\rm line}$, and modeled line profiles shaped by scale-height-dependent turbulence as Lorentzian profiles. They obtained distances $R$ of the line emitting region using reverberation mapping and modeled the scale height $H$ according to
\begin{equation}
    H/R = (1/\alpha)(v_{\rm turb}/v_{\rm rot})
    \label{eq:scaleheight}
,\end{equation}
given by \citet{pringle81}, where $\alpha$ is the viscosity parameter of order unity. The velocities give the turbulent and rotational velocites, respectively. \citet{kollatschny13a} and \citet{kollatschny13b} found a flattened distribution for the \Hb{} emitting region, and in particular that the height of the line emitting region increases with distance to the SMBH, resulting in a flat bowl-shaped geometry. Strikingly, \citet{kollatschny13a} found for a large sample of AGN that each emission line was associated with a specific turbulent velocity, and in particular that the turbulent component in \Hb{} profiles corresponded to turbulent motions of $v_{\rm turb} \sim 400$\,\kms{}. \citet{goad12} found that Lorentzian profiles of the emission lines naturally arise for low-inclination systems in their bowl-shaped BLR model, and evidence for bowl-shaped BLR geometries \citep[e.g.,][and references therein]{ramolla18} or in general flattened or disk-like geometries \citep[e.g.,][]{gaskell07,pancoast14, williams18} have also been found by other studies.

As the temporal sampling of the presented variability campaign of \ngc{} during the transient event is too low to perform classical reverberation mapping, we are not able to obtain geometrical information from the transfer function. Therefore, we cannot make a precise prediction about the exact shape of the BLR; that is, whether it is rather disk-like or bowl-shaped. However, as the aforementioned models associate turbulence with scale height above the disk, we deduce a vertical stratification of the BLR; that is, \Hb{} is emitted at greater scale height than \ion{Ca}{ii} and \ion{O}{i}:  For $\alpha$ set to 1, the turbulent velocities  $v({\rm CaII})_{\rm turb} = 200$\,\kms{} (see \ref{sec:disk_origin_discussion}) and $v({\rm H\beta})_{\rm turb} = 900$\,\kms{} as well as FWHM(CaII) $=1920$\,\kms{} and FWHM(H$\beta$) $=2200$\,\kms{}, we obtain an $H/R$ value of $\sim 0.1$ and $\sim 0.4$ for \ion{Ca}{ii} and \Hb{}, respectively. Therefore, \Hb{} is being emitted roughly at four times the height of \ion{Ca}{ii}.

Although the \ion{He}{i} profiles do not allow for a reliable fit with the elliptical disk model alone, we note that we can in principle construct such ``smeared out'' broader profiles (in comparison to \Hb{}) by decreasing the inner and outer pericenter distance of the emitting region while at the same time increasing the internal broadening. This indicates that the \ion{He}{i} emission is generated at shorter distances from the ionizing continuum (``further in''), paired with higher internal broadening, which can be interpreted as turbulence due to greater scale height and therefore a higher layer in the disk. This is in agreement with results of \citet{kollatschny13b}, who showed that high-ionization lines are emitted closer to the SMBH, but at greater heights.

In conclusion, we term the BLR in \ngc{} to be \textit{disk-dominated}. This means that all broad emission lines analyzed in this study (\Hb{}, \ion{He}{i}, \ion{Ca}{ii}, and \ion{O}{i}) show the kinematic signature of an elliptical accretion disk. While this signature is clearly visible in the line profiles of \ion{Ca}{ii}, and \ion{O}{i} as they are being emitted closest to the midplane of the disk, it is smeared out by turbulence (and in some part by the presence of additional emission components) in the case of \Hb{} and \ion{He}{i}.

%
\subsubsection{Temporal evolution of the Balmer profiles: A possible disk wind.}\label{sec:Balmer_profile_variations_discussion}
%
We show in \ref{sec:optical_profiles_results} and \ref{sec:Bowen_fluorescence_discussion}, that \Hb{} and \ion{O}{i}$\,\lambda8446$ exhibit blueward drifts by $-240$\,\kms{} and $ - (150 - 200)$\,\kms{}, respectively, during the rising phase of the transient event (between Spectrum 1 and Spectrum 2). When fitting line profiles with accretion disk models, velocity shifts due to other physical processes affect the (a)symmetry of the profile and may prevent the determination of the true disk parameters if not taken into account. This problem regarding velocity shifted line profiles has already been addressed by \citet{eracleous95} and \citet{eracleous03} with respect to possible orbital motions in a binary system used to explain the formation of eccentric disks. In the following, we investigate the impact of  the blueward drift on the best-fit parameter set and discuss possible explanations for the observed line profile shifts.

Supposing that \ion{Ca}{ii} exhibits a velocity shift similar to \ion{O}{i}$\,\lambda8446$, and that the line profile does not change significantly (as indicated by the almost identical \Hb{} profiles during the onset of the transient event), we perform a fit on a \ion{Ca}{ii}$\,\lambda8662$ profile that we artificially redshift by $+200$\,\kms{}. This way, we compensate the blueward drift of the line during the transient event and obtain, presumably, the original \ion{Ca}{ii}$\,\lambda8662$ before the transient event. We find that we are able to fit the shifted \ion{Ca}{ii}$\,\lambda8662$ profile with the disk parameters $\xi_1 = (1140 \pm 800) $\,r$_g$, $\xi_2 = (4468 \pm 1600)$\,r$_g$, $i = (5.64 \pm 3.00)^{\circ}$, $\phi_0 = (207.94 \pm 26.00)^{\circ}$, $\sigma =(85 \pm 10)$\,\kms{}, $e = (0.55 \pm 0.40)$, and $q = (4.50 \pm 0.90)$. This fit largely reproduces the best-fit parameters obtained in \ref{sec:fitting_results} within their error margins, except for the inner pericenter distance $\xi_1$, which decreased by half and is now slightly below the determined margin of error.

We now discuss the blueward drift of the \Hb{} profile in the context of line profiles presented in previous studies. The \Hb{} profile shifted by $ -240$\,\kms{} within $\sim 750$\,days (between Spectrum 1 and Spectrum 2) and by additional $ -130$\,\kms{} within the next $270$\,days (between Spectrum 2 and Spectrum 3; see Fig.~\ref{fig:Hbeta_blueward_drift}). However, previous low-state observations over the last 30 years indicate a constant redshift of the \Hb{} line. \citet{kriss91} found an H$\beta_{\rm broad}$ velocity shift of $\sim +700$\kms{} in their HST/FOS spectrum from 1991. Likewise, we find an \Hb{} redshift of $(+670 \pm 50)$\,\kms{} in the mean spectrum of a variability campaign on \ngc{} we performed from 2012 November to 2013 March from which we presented a spectrum in \citet{ochmann20}. A velocity shift of $(+375 \pm 6$)\,\kms{} in a GMOS spectrum obtained in 2013 October was reported by \citet{silva18}. However, their analysis is primarily based on \Ha{} due to its preferable S/N, and the Gaussian fit to H$\beta_{\rm broad}$ does not capture the line profile of \Hb{} well. As the \Hb{} profile from \citet{silva18} and the line profile from our 2012 to 2013 campaign are, by visual inspection, virtually identical, we conclude that the velocity shift in the GMOS spectrum from 2013 October is in fact closer to $\sim +700$\,\kms{}. The stability of the \Hb{} (and \Ha{}) profile was already pointed out by \citet{alloin85}, who noticed that no considerable changes in the shape of the broad components were present, despite ample changes in total flux during their observations lasting from 1980 to 1982.

Given the stability of the line profile on timescales of years, the consistent velocity shift of $\sim + 700$\,\kms{} for H$\beta_{\rm broad}$ over the last 30 years, and indications for slightly different velocity shifts in \Hb{} and \ion{O}{i}$\,\lambda8446$, a disk wind might serve as an intuitive explanation for the observed blueward drift of the \Hb{} profile.
Accretion disk winds are thought to play an important role in the formation of the BLR \citep{czerny11}, and they have been invoked to explain several observational key features of the BLR, for example, the lower redshifts of low-ionization broad emission lines (BELs) in comparison to high-ionization BELs \citep{gaskell82}, or the broad absorption lines (BALs) seen in the UV spectra of some AGN \citep{weymann92}. \citet{gaskell82} tied the different redshifts between low- and high-ionization BELs to two different ionization zones, and it was later suggested that these zones are the low-ionization surface layer of a high-density accretion disk and a high-ionization, low-density wind launched from the accretion disk \citep[][and references therein]{collin-souffrin86, collin-souffrin88}. Generally, accretion disk winds can have a severe impact on the observed line profile shapes, in particular depending on the optical depth of the wind \citep[e.g.,][]{chiang96, murray97, flohic12}. If the optical depth is low, accretion disk winds can induce blueshifts on lines without changing their profiles \citep[][]{eracleous03}. A disk wind would be consistent with results from \citet{parker19}, who found indications for a $v \sim 500$\,\kms{} outflow being launched during the outburst in \ngc{}. On basis of the results presented in this paper, namely the blueward drift of the line profiles and differing velocity shifts between individual lines, we propose a launching accretion disk wind as an explanation for the blueward drift of \Hb{} and \ion{O}{i}\,$\lambda8446$ during the rising phase of the transient event in \ngc{}.

We note that changes in the line profiles caused by accretion disk winds might explain the more complex \Hb{} profiles observed in later stages of the transient event (see Fig.~\ref{fig:clean_broad_Balmer_profiles}), when \ngc{} transitions from a Sy 1.2 back to a Sy 1.8 galaxy \citep{elitzur14}.

%
\subsubsection{Evidence for BLR (and AD) inhomogeneities}\label{sec:disk_inhomogeneities_discussion}
%
We show in \ref{sec:robustness_line_profile_discussion} that the small-scale features in the \ion{O}{i} and \ion{Ca}{ii} line profiles are not caused by absorption or blending effects, but are genuine features of the double-peaked profiles. The homogeneous elliptical disk model applied in \ref{sec:fitting_results} is not able to account for these features, and the small-scale variations in the profile remain imprinted on the residual flux. The most prominent deviation from the elliptical-disk profile is the trident structure in the red peak, that appears to be constituted by three individual, overlapping peaks. This might, for example, indicate that the inner orbit of the BLR disk in \ngc{} is in fact split, such that the high-velocity part of the profile splits up. Although our current modeling cannot distinguish if the small-scale line features in the two-component BLR model originate from the disk or the gas above the disk plane, the trident in the red peak makes it intriguing to associate the small-scale deviations with inhomogeneities in the BLR disk structure, which we argue to be tightly connected to an underlying accretion flow. Inhomogeneities appear to be a key feature in AGN accretion disks \citep[e.g.][]{dexter11, jiang2020}, and are, for example, invoked to explain the variability in AGN. Inhomogeneities in the BLR have been invoked in the literature in the past to explain broad-line profile changes \citep[e.g.,][]{veilleux91,wanders95} on timescales of years, and more recently  \citet{horne21} identified azimuthal structures rotating in the BLR of NGC\,5548, revealing that the BLR is not a homogeneous entity. Structural BLR models incorporating inhomogeneities in the BLR invoke, among others, spiral arms or hot spots in the disk \citep[e.g.,][]{gilbert99,storchi-bergmann03, jovanovic10} and spiral arm models are able to explain the generally observed differences between average and rms profiles as well as complex subfeatures in velocity-delay maps \citep{wang22, du23}. The detection of BLR inhomogeneities in the profiles of \ion{O}{i} and \ion{Ca}{ii} in \ngc{} suggests that reverberation mapping of these lines could be a powerful instrument to investigate the BLR structure in AGN in more detail. This has already been noted by \citet{matsuoka08}.

%
\subsubsection{A turbulent and disk-dominated BLR model as an explanation for the observed emission line profiles}\label{sec:turbulent_disk_BLR_discussion}
\begin{figure}[h!]
    \centering
    \includegraphics[width=0.47\textwidth,angle=0]{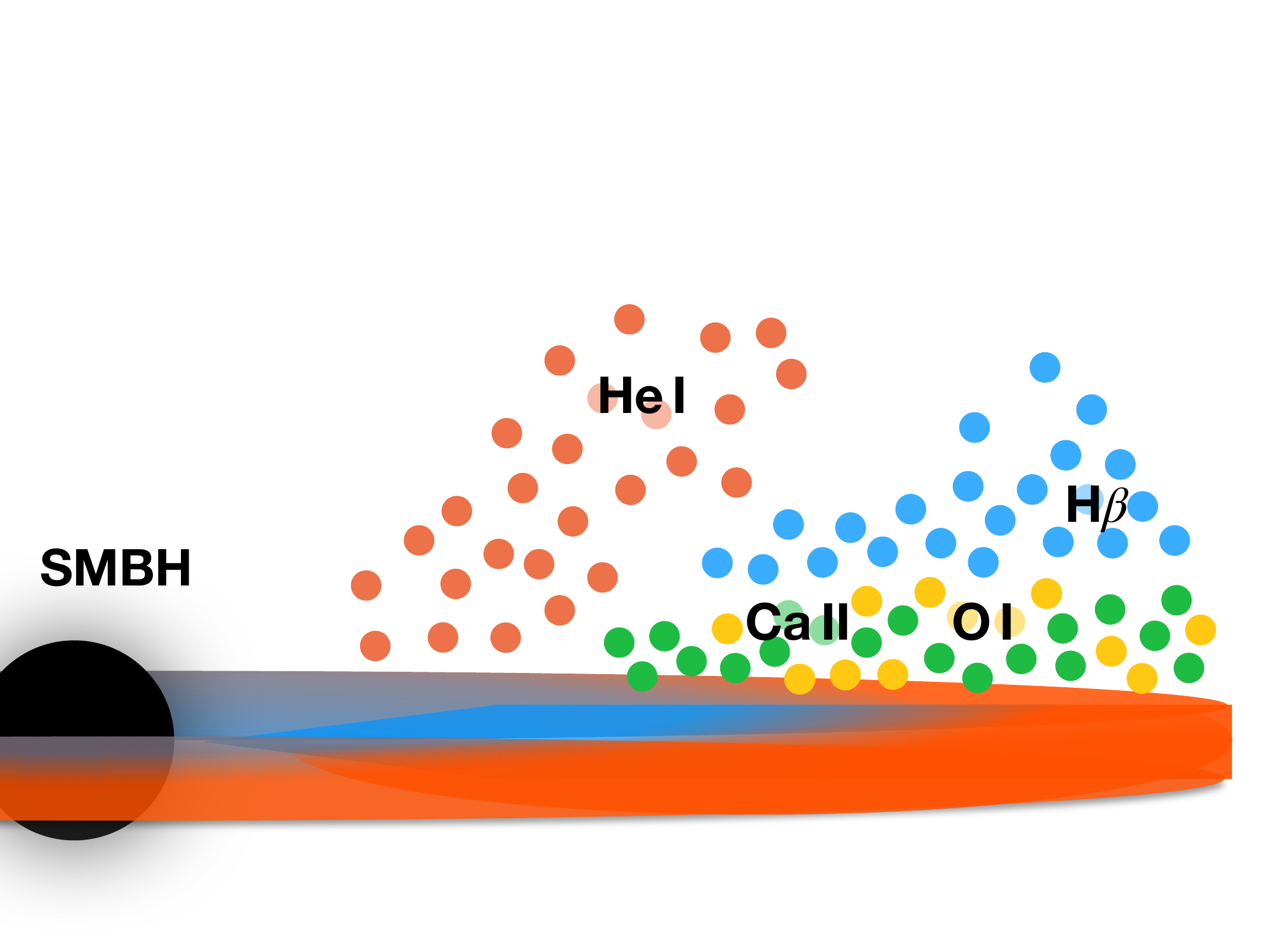}
    \caption{Qualitative edge-on view of the BLR geometry suggested for \ngc{} (not to scale and the height of the BLR above the disk plane is largely exaggerated for clarity).  The blue and red color of the disk vaguely represents the transition from a hotter inner to a cooler outer accretion disk. The BLR has a flattened distribution, with \ion{O}{i} and \ion{Ca}{ii} being emitted close to the midplane of the accretion disk. \Hb{} is emitted at greater scale height. \ion{He}{i} is emitted at shorter radial distance from the ionizing source.} 
    \label{fig:BLR_sketch}
\end{figure}

%
We show above that the line profiles in \ngc{} during the transient event are shaped by the superposition of at least three effects: (1) The asymmetric and double-peaked \ion{O}{i} and \ion{Ca}{ii} profiles  can be modeled by an elliptical, eccentric disk component; (2) the \Hb{} line profiles can be modeled as broadened versions of these profiles, a fact that we ascribe to scale-height-dependent turbulence; and (3) the line profiles exhibit blueward drifts during the rising phase of the transient event, a circumstance that might be interpreted as the signature of a low-optical-depth disk wind being launched during the transient event. In the emerging BLR picture, the BLR is not only radially stratified with regard to its ionization structure, but also vertically stratified with regard to scale-height-dependent turbulence. Based on the larger FWHM of the \ion{He}{i} lines in comparison to \Hb{}, \ion{O}{i}, and \ion{Ca}{ii}, \ion{He}{i} is presumably emitted at shorter radial distance from the ionizing source than the other lines. Furthermore, according to our model \Hb{} is emitted at greater scale height than \ion{O}{i}, and \ion{Ca}{ii}, but roughly at the same radial distance. To illustrate, we show a qualitative sketch of our BLR model in Fig.~\ref{fig:BLR_sketch}.

%


%
\section{Conclusions}\label{sec:conclusions}
%
In this study, we present results of a spectroscopic variability campaign performed on \ngc{} during its transient event from 2017 to 2019. In particular, we analyzed the emission line profiles of \ion{Ca}{ii}, \ion{O}{i}, \ion{He}{i}, \Hb{}, and \Ha{}, and the corresponding profile changes during the event. Our results can be summarized as follows: 

\begin{enumerate}
    \item \ngc{} exhibits strong spectral changes during its transient event from 2017 to 2019. We observe the emergence and subsequent fading of a strong, power-law-like blue continuum as well as the brightening and subsequent fading of strong Balmer, \ion{He}{i}, and \ion{He}{ii} lines. In addition, we observe variable coronal line emission of [\ion{Fe}{vii}], [\ion{Fe}{X}], and [\ion{Fe}{XI}]. Further, more detailed results on the spectral variability in \ngc{} during its transient event will be presented in future publications.
    
    \item We report the detection of double-peaked profiles of the Bowen-fluorescence line \ion{O}{i}$\,\lambda8446$ and \ion{Ca}{ii} with redward asymmetry during the transient event. To our knowledge, this is the first time that genuine double-peaked \ion{O}{i}$\,\lambda 8446$ and \ion{Ca}{ii} triplet emission line profiles in AGN have been presented in the literature. Moreover, we find indications that \ion{O}{i}\,$\lambda8446$ as well as NIR \ion{Ca}{ii} triplet emission was already present before the onset of the transient event.
    
    The \ion{O}{i}$\,\lambda 8446$ and \ion{Ca}{ii} triplet emission line profiles are well approximated by a combination of emission from an eccentric disk with a low inclination angle of  $i = (8.10 \pm 3.00)^{\circ}$ and an emission component from BLR gas above the disk. In addition, we interpret small-scale features in the \ion{Ca}{ii} line profile as evidence for inhomogeneities in the broad-line region associated with the accretion disk.
    
    \item We find a NIR \ion{Ca}{ii} triplet ratio of 1:1:1 and an \ion{O}{i}$\,\lambda8446$-to-\ion{Ca}{ii}$\,\lambda8662$ ratio of 0.85:1. Based on the results of previous studies by other authors, this suggests that the emission of \ion{O}{i} and \ion{Ca}{ii} originates from approximately the same region, namely a region of high (column) density with the kinematic signature of a rotating disk.
    
    \item Based on the stable line profiles and the constant FWHM of \Hb{} during the event, and $L\beta$-pumped \ion{O}{i}$\,\lambda8446$ emission already being present before the brightening in \ngc{}, we conclude that the broad-line region associated with the disk was already present before the beginning of the transient event. More specifically, we postulate that the elliptical disk and the associated broad-line region were not formed during the transient event, but were present before the event began. 
    
    \item We show that the Balmer line profiles in \ngc{} exhibit the same redward asymmetry as the \ion{O}{i} and \ion{Ca}{ii} lines, and that, in comparison, they have a slightly larger FWHM. We are able to reconstruct the \Hb{} and \Ha{} rms profiles by convolving the double-peaked \ion{Ca}{ii}$\,\lambda8662$ profile with a Lorentzian with a  width of $\sim 900$\,\kms{}. We suggest that this might be interpreted as broadening due to turbulent motions of the Balmer-line-emitting gas and that this turbulence smears out the double-peaked profile. According to this interpretation, the region of Balmer line emission in the broad-line region is most likely situated at greater scale height above the accretion disk than the region emitting \ion{Ca}{ii}.
    
    \item We find indications for a blueward drift of the total \Hb{} and \ion{O}{i}\,$\lambda8446$ profiles during the rising phase of the transient event. These varying velocity shifts might be interpreted as the emergence of a low-optical-depth wind being launched during the transient event. Except for this launching disk wind, the observed kinematics of the BLR remain largely unchanged during the transient event.
\end{enumerate}
Our results suggest a flattened (disk-dominated) broad-line region geometry, with a vertical stratification due to winds in combination with turbulence.
Our results demonstrate the usefulness of spectroscopic transient variability campaigns with high data quality, even when sampling is sparse. We especially point out the usefulness of observations of \ion{O}{i}$\,\lambda8446$ and the NIR \ion{Ca}{ii} triplet, which is due to the close proximity of their emission region to the accretion disk. Further, more densely sampled spectroscopic observations of transient events are strongly desired.

\begin{acknowledgements}
The authors thank the anonymous referee for detailed and useful comments, which contributed significantly to the improvement of the manuscript. Furthermore, the authors thank Peter~M. Weilbacher for helpful discussions regarding the MUSE reduction pipeline, Dragana Ili\'c for enlightening discussions about emission lines in optical spectra, Paul~I. Schwarz for computational advice as well as helpful discussions, and Bj\"orn M\"uller for discussions regarding mathematical details of the fit to the \ion{Ca}{ii} profile. The authors greatly acknowledge support by the DFG grants KO 857/35-1 and CH 71/34-3. MWO gratefully acknowledges the support of the German Aerospace Center (DLR) within the framework of the ``Verbundforschung Astronomie und Astrophysik'' through grant 50OR2305 with funds from the German Federal Ministry for Economic Affairs and Climate Action (BMWK). ERC acknowledges support from the National Research Foundation of South Africa.

Five spectra reported in this paper were obtained with the Southern African Large Telescope (SALT under proposal codes 2018-1-DDT-004, 2018-1-DDT-008 (PI: Kollatschny) and 2018-2-LSP-001 (PI: Buckley). In addition, this paper is based on observations made with ESO Telescopes at the La Silla Paranal Observatory under programme IDs 096.D-0263 and 0100.B-0116, obtained from the ESO Science Archive Facility.

\end{acknowledgements}


\bibliographystyle{aa} 
\bibliography{literature} 

\end{document}